# Timestamp-Aware Spatio-Temporal Graph Contrastive Learning for Network Intrusion Detection★

Jianli Dai[a,1], Guangwei Wu[a,*], Jiacheng Li[a], Weiping Wang[b], An He[a] and Xinjun Xiao[a]

[a]Central South University of Forestry and Technology, School of Computer Science and Mathematics, Changsha, 410004, Hunan, P.R. China
[b]Central South University, School of Computer Science and Engineering, Changsha, 410083, Hunan, P.R. China

ABSTRACT

Given their effectiveness in modeling the relational structure among network traffic flows, graph neural networks (GNNs) have been widely adopted in network intrusion detection systems (NIDSs). However, most existing GNN-based NIDS approaches focus on the relational structure of traffic flows, and treat them as temporally independent, which limits their ability to cope with evolving attack behaviors. Moreover, their reliance on supervised or semi-supervised learning often restricts generalization to unseen attacks. To address these limitations, we propose a novel self-supervised GNN-based framework. To the best of our knowledge, the proposed model is among the first self-supervised GNN-based NIDS models to explicitly leverage real timestamps, which provides faithful temporal dependencies for representation learning. We first construct a series of temporal graphs from network traffic flows according to their timestamps, and then employ an E-GraphSAGE and LSTM based encoder to fully extract temporal information and spatial dependencies of network traffic, without introducing time-costly attention mechanisms. A multi-view graph contrastive learning (GCL) scheme is introduced, where temporal, spatial, and feature contrasts are jointly performed to capture temporal continuity, preserve structural consistency, and improve the generalization and robustness of the learned representations, respectively. In addition, a gradient-norm-based adaptive weighting strategy is designed to optimize the contrastive loss weights. Experimental results on four representative NIDS datasets with real timestamps demonstrate that our method significantly outperforms existing self-supervised approaches and achieves performance comparable to the supervised state-of-the-art GNN method, while maintaining high computational efficiency. These results further validate the effectiveness of explicitly modeling temporal dependencies in NIDSs. The code and resources of our experiments are available at https://github.com/Rory6235/STG-NIDS.

## 1. Introduction

The exponential development of the Internet and Internet of Things (IoT) technologies has led to the widespread deployment of networks, which consist of heterogeneous devices (e.g., computers, sensors, and smart devices) and interconnected communication components, making them an essential part of modern society [1]. Massive volumes of network traffic flows are continuously generated by host-to-host communications in heterogeneous networks and evolve over time, exhibiting complex temporal and spatial dependencies [2]. Such complex and evolving characteristics expand the attack surface of networks, resulting in ubiquitous cyberattacks that threaten confidentiality, integrity, and availability [3]. In particular, sophisticated attacks such as denial-of-service (DoS), distributed denial-of-service (DDoS), and scanning attacks frequently exhibit temporal and spatial correlations among multiple hosts, which pose significant challenges to accurately identifying malicious activities. To address these challenges, numerous security technologies have been developed, among which network intrusion detection systems (NIDSs) play a crucial role in monitoring network traffic flows and identifying malicious activities in a timely and automated manner [4].

The prevailing solutions in NIDS are primarily driven by machine learning (ML) and deep learning (DL) [5]. Although traditional ML-based NIDS performs effectively on flat and handcrafted network traffic features, it struggles to capture the structural dependencies and interactions among network traffic flows [6]. In contrast, DL-based models are capable of learning non-linear, high dimensional, and sequential patterns from raw network traffic flows, thereby significantly enhancing the detection capability of complex and evolving attacks [7]. Network traffic flows in NIDS can be naturally modeled as a graph, where hosts are represented as nodes, and the network traffic flows exchanged between hosts are represented as edges. As a subfield of deep learning, graph neural networks (GNNs) have been demonstrated great ability in exploiting graph-structured data, enabling effective modeling of dependencies and interactions between hosts [4]. Nevertheless, representative GNN-based methods, including GraphSAGE and graph attention networks (GATs) [8], primarily adopt the node-centric learning paradigm and directly rely on node features, thus ignoring edge features that characterize network traffic behavior. This node-centric design is inherently limited for NIDS, as most discriminative information, such as durations, timestamps, and protocol, are carried by network traffic flows rather than by hosts. To address this limitation, several studies [9] have extended GNN models to jointly exploit node and edge features, allowing the models to more effectively characterize network traffic behaviors. By integrating node and edge features, these approaches can capture the intricate dependencies among hosts and flows, achieving excellent performance in network intrusion detection tasks.

★This work is supported by the National Key Research and Development Program of China under Grant 2023YFB3106900, the Key Scientific Research Projects of Hunan Provincial Education Department under Grant 25A0266 and 24A0197.

*Corresponding author
guangweiwu@csuft.edu.cn (G. Wu)

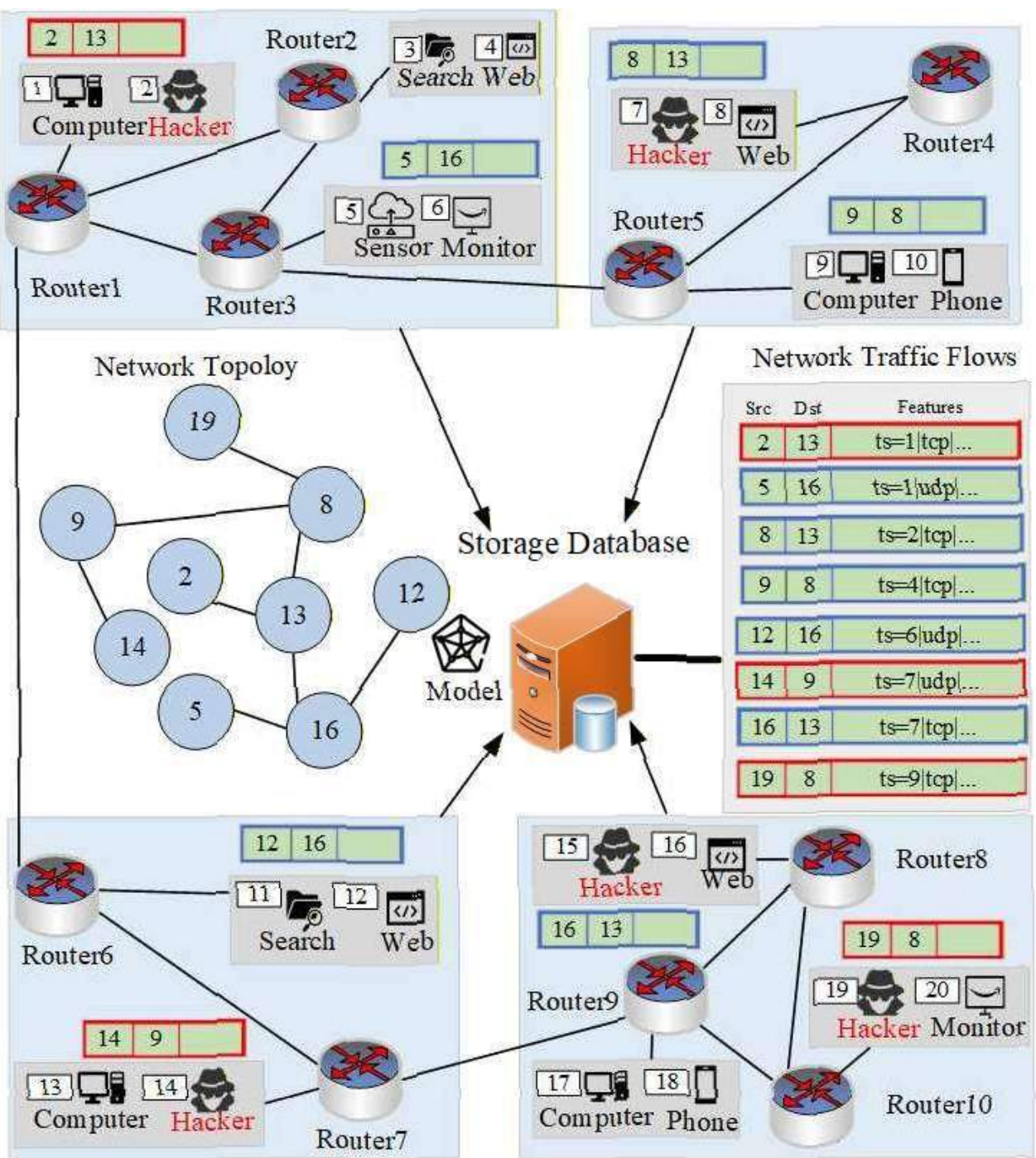

**Figure 1:** Deployment overview of the proposed framework. The table on the center-right lists network traffic flows collected by storage database within a specific time interval, and the graph on the center-left depicts the spatial topology constructed from these flows. We mention that the spatial topology of the network is often inconsistent with its underlying physical connectivity.

Most existing GNN-based models for NIDS are trained in a supervised or semi-supervised manner [9, 10], which heavily depends on large-scale labeled data. Such methods require labor-intensive manual annotation of network traffic flows, making the training process costly and time-consuming. Moreover, the complex and dynamically evolving nature of network traffic flows often leads to previously unseen attack patterns, which further limits the generalizability of supervised models [11]. To alleviate the dependence on expensive labeled data, self-supervised learning has emerged as a promising paradigm, enabling GNN models to learn informative representations from unlabeled graph-structured data [12]. Recent studies have introduced self-supervised learning and graph contrastive learning (GCL) into NIDS, learning discriminative representations of network traffic flows by contrasting different graph views without relying on labeled annotations [13, 14, 15]. Caville et al. [14] proposed Anomal-E, a self-supervised GNN-based NIDS that explicitly incorporates edge features and topology. Xu et al. [15] proposed NEGAT+NEGSC that adopts attention-based multi-perspective contrast to enhance both binary and multiclass classifications in NIDS.

Given that real-world network environments evolve continuously, network intrusion detection inherently involves both spatial and temporal characteristics. Malicious activities often present distinct spatial structures such as botnet communications, lateral movement patterns, abnormal connectivity while simultaneously exhibiting temporal patterns including DoS/DDoS bursts, progressive infiltration. However, most existing self-supervised GNN-based NIDS built on static graphs tend to overlook the inherent temporal dynamics of network traffic flows. Spatio-temporal GNNs provide an effective way to jointly model spatial dependencies and temporal representations from graph-structured data [16] and therefore have been increasingly adopted for NIDS [17, 18, 19]. Langendonck et al. [17] employed self-supervised pre-training and sliding-window temporal graphs to enable near real-time intrusion detection. Escriche et al. [18] proposed CCSTGN, a channel-centric spatio-temporal GNN that models communication channels as nodes with continuous temporal representations. Wu et. al [19] further proposed TCG-IDS to exploit temporal information through self-supervised contrastive with generated timestamps.

To address the aforementioned limitations, we propose a novel spatio-temporal self-supervised GNN-based framework for network intrusion detection that first explicitly incorporates real timestamps to model the temporal evolution of network traffic flows. The overview architecture of the proposed framework is illustrated in Fig. 1. Unlike existing GNN-based NIDS that rely on labeled data or artificially generated timestamps, our framework adopts self-supervised learning to leverage spatio-temporal graphs to jointly capture spatial structures and temporal dynamics of network traffic flows.

The main contributions of this paper are summarized as follows:

- We propose a novel self-supervised GNN-based framework for NIDS. To the best of our knowledge, the proposed framework is among the first to explicitly model the intrinsic temporal dependencies among network traffic flows using the dataset-provided real timestamps. By deeply integrating temporal and spatial information via a spatio-temporal encoder and a timestamp-aware contrastive objective, the proposed model fully captures latent spatio-temporal patterns while maintaining efficient inference without relying on attention mechanisms.

- We design three self-supervised graph contrastive objectives to capture latent spatio-temporal representations and alleviate label dependence under evovling attack patterns. Temporal contrast aligns a node's predicted representation with its corresponding representation at the next time interval to maintain temporal smoothness. Spatial contrast increases the similarity between the representations of each node and its neighbors within the same temporal graph to preserve local structural coherence. Feature contrast constructs augmented and perturbed samples of nodes based on spatio-temporal weights to improve the model's generalization and robustness. To coordinate these objectives, we further propose a gradient-norm-based adaptive weighting strategy to dynamically balance multiple contrastive losses.

- We conduct extensive evaluations on four widely used datasets with real timestamps from both theoretical and experimental perspectives. Theoretical time complexity analysis shows that replacing edge-dependent attention mechanisms with explicit temporal dependency modeling yields a more efficient computational scheme. Experimental results show that our model consistently outperforms related representative self-supervised models and state-of-the-art GNN-based supervised method on both binary and multiclass classification tasks, while achieving higher computational efficiency. Moreover, the ablation studies illustrate the importance of temporal information in models for NIDSs.

# 2. Related Works

The section reviews recent studies from three aspects: network intrusion detection system, graph self-supervised learning, and spatio-temporal graph neural networks. The review aims to clarify their strengths and limitations, and to establish the motivation for our proposed approach.

## 2.1. Network Intrusion Detection System

Network intrusion detection system (NIDS) is a security mechanism designed to monitor and analyze network traffic flows in real time for detecting malicious activities.

Recently, machine learning (ML) and deep learning (DL) techniques have been widely adopted in NIDS due to their strong capability in pattern recognition and data-driven modeling. ML-based NIDS typically exploys supervised learning methods to classify network traffic flows and identify malicious activities. Lee et al. [20] applied an extreme learning machine (ELM) with equation constrained optimization to NIDS and proposed an adaptive incremental learning method to determine the optimal number of hidden neurons. Compared with traditional ML-based approaches, DL-based NIDS exhibit significant advantages in modeling complex, high-dimensional network traffic flows and enabling automated feature extraction, which is critical for behavior and anomaly analysis. Vinayakumar et al. [21] systematically investigated the application of convolutional neural networks (CNNs), recurrent neural networks (RNNs), long- and short-term memory (LSTM) and gate recurrent units (GRUs) in NIDS. Experiment results demonstrate that CNN-based models and their variants consistently outperform traditional ML-based methods in network attack detection, among which the CNN-LSTM architecture achieves the best performance. Furthermore, Haitao et al. [22] proposed a multimodal sequential NIDS by integrating deep autoencoders with LSTM networks to jointly capture structural information and temporal dependencies shared across similar network connections. Owing to their strong capability in capturing structural dependencies and interaction patterns among network hosts, graph neural networks (GNNs) have attracted increasing attention in NIDS. Consequently, GNN methods have been widely explored by researchers for NIDS applications. Lo et al. [9] proposed E-GraphSAGE, a GNN-based NIDS model that incorporates edge features and graph topology to effectively model network traffic flows, enabling efficient detection of complex attacks in IoT networks and achieving superior detection performance compared to existing models.

However, most existing models for NIDS are based on supervised or semi-supervised learning paradigms and rely heavily on labeled data, which may limit the generalization in different network environments.

## 2.2. Graph Self-supervised Learning

Self-supervised learning (SSL) has gained significant attention due to its ability to learn robust representations without requiring labeled data. Existing SSL methods can be broadly categorized into generative-based and contrastive-based approaches. Generative SSL aims to reconstruct or model the underlying data distribution in a task-agnostic manner, enabling its applicability to various downstream tasks such as classification and generation [23]. In contrast, contrastive SSL focuses on learning disciminative representations by maximizing the agreement between positive pairs while distinguishing negative samples, and has demonstrated superior performance in many representation learning tasks. As a prominent contrastive SSL paradigm, graph contrastive learning (GCL) reduces the dependence on labeled data by exploring the inherent similarities and differences among graph components, such as nodes, subgraphs and entire graphs.

The origins of graph self-supervised learning can be traced back to early studies on unsupervised graph embedding methods [24, 25], which learn node representations by maximizing the agreement between contextual nodes within truncated random walks [11]. Subsequently, graph autoencoders (GAEs) [26], were proposed as a widely adopted unsupervised learning framework that reconstructs graph structures and can be viewed as an early form of graph self-supervised learning. Furthermore, deep graph infomax (DGI) [27] introduced a mutual information maximization objective between node-level and graph-level representations as a self-supervised proxy task, marking an important milestone in the development of graph self-supervised learning. Y. You et al.[28] introduced an effective self-supervised learning framework, GCA, which leverages graph augmentations to learn roboust representations. Generative Subgraph Contrast (GSC) [29] introduced an adaptive subgraph generation strategy within a contrastive learning framework to learn efficient and robust representations, where the optimal transmission distance is employed as a similarity measure between subgraphs.

In NIDS, Cavillea et al. [14] proposed Anomal-E, a self-supervised intrusion detection framework that exploits both edge features and graph structural information achieving great results in binary classification tasks. Xu et al. [15] proposed an improved self-supervised method, NEGSC, which contrasts subgraphs generated from a center node and its neighboring nodes, and demonstrates superior performance in both binary and multiclass intrusion detection. C. Wu et al. [19] designed TCG-IDS, a self-supervised IDS framework that leverages temporal contrastive graph neural network (GNN) and multiple contrasting strategies with time dependency, achieving strong performance on NetFlow datasets. However, because the dataset used in TCG-IDS does not provide real timestamps for network flows, its temporal graphs are constructed from a generated timeline constructed from flow duration, rather than from dataset-provided real timestamps. Such a generated timeline may not faithfully reflect the real temporal order and evolution patterns of network traffic.

In conclusion, most existing self-supervised NIDS methods do not explicitly model temporal information and treat timestamps as regular features, making it fail to capture the temporal continuity of host behaviors and the evolution of attacks over time. The only method [19] that incorporates temporal dependencies is designed based on generated timestamps, which is constructed using the duration of network flows rather than the real timestamps. This clearly prevents the method from preserving the true ordering of network traffic flows, thereby limiting its adaptability to dynamic real-world scenarios.

### 2.3. Spatio-Temporal Graph Neural Networks

Spatio-temporal graph neural networks (STGNNs) constitute a deep learning framework that integrates GNNs with time-series modeling techniques to capture both spatial dependencies and temporal dynamics in complex data. The core idea of STGNNs is to represent spatio-temporal data as graph-structured information, where nodes correspond to network hosts and edges as network traffic flows between network hosts, which are constructed according to the temporal evolution of network traffic flows. By jointly modeling graph topology and temporal evolution, STGNNs learn expressive spatio-temporal representations through GNNs.

In recent years, STGNNs have been extensively studied across a wide range of Application domains, including network intrusion detection, anomaly detection, and traffic forcasting. Rossi et al. [30] introduced TGN, a general and efficient deep learning framework for dynamic graphs represented as sequences of time events, which integrates memory modules and graph-based operators while requiring substantial memory during training. Chen et al. [31] proposed DySubC, a self-supervised dynamic learning framework which contrasts temporal subgraphs by introducing a novel temporal subgraph sampling method, enabling the model to jointly capture the structural and evolutionary features of dynamic graphs. H. Wang et al. [32] proposed an intelligent digital twin framework for IoT attack detection, which selects informative feature subsets based on information gain and employs a simplified CNN model to capture the temporal patterns of attack behaviors. B. Yu et al. [33] formulated time series prediction as a graph problem and introduced a spatio-temporal graph convolutional network (STGCN) with fully convolutional structures, enabling efficient modeling and faster training. Duan et al. [34] developed a network anomaly detection framework based on continuous temporal graph (CTG) neural network, which refines the specific information interactions, thus naturally incorporating new node access behaviors into the feature extraction of CTG neural networks.

These studies collectively highlight the importance of jointly modeling spatial dependencies and temporal information, enabling models to improve detection accuracy and robustness in diverse network environments. However, most existing models are either developed for general dynamic graph learning or tailored for some special time series prediction. As a result, they are hard to applied directly to NIDS in evolving network environments, where the constructed graph in network has unique characteristics. Specifically, most of the information is often carried by edges, and the graphs tend to be dense and multigraph.

## 3. Methodology

In this section, we first describe the data pre-processing and the feature selection engineering, and present the notations of our approach in detail. Then, a spatio-temporal self-supervised GNN-based framework for network intrusion detection that first leverages real timestamps is introduced. which is shown in Fig. 2. The overall framework consists of three main components: temporal graph construction, spatio-temporal GNN encoder, and graph contrastive learning. Specifically,

1) Temporal graph construction aims to divide the network flows into multiple temporal graphs in the order of real timestamps.

2) Spatio-temporal GNN encoder captures spatial dependencies and temporal information and learn diverse high-dimension graph representation, which consists of E-GraphSAGE used for aggregating edge features, fusion layer to combine the temporal and spatial information, and LSTM to capture the hidden states and cell memories.

3) Graph contrastive learning adopts three strategies to guide the model in learning latent embeddings by taking advantage of temporal, spatial and feature similarities within each temporal graph.

Finally, a detailed time complexity analysis is presented for our model and the compared models.

### 3.1. Data Pre-processing

The preprocessing data is crucial to ensure data consistency and improve model performance. The steps in our study are as follows: data cleaning, categorical feature coding and data normalization. Initially, raw network traffic data, including features like source/destination IP addresses, ports, packet size and flow information, are structured into a tabular format, where missing or erroneous values, such as NaN or undefined symbols (e.g., '-') are removed. Subsequently, categorical features are encoded to numerical values using LabelEncoder, which assigns a unique integer to each category. For a categorical feature with unique values, LabelEncoder would map each category to an integer. Ultimately, the numerical values of features are standardized by StandardScaler, which transforms the data to have a mean of 0 and a standard deviation of 1. This process can ensure that each feature contributes equally to the model learning process. It is worth noting that in our method, time-relevant features, such as timestamps, are selected as separate columns for subsequent use.

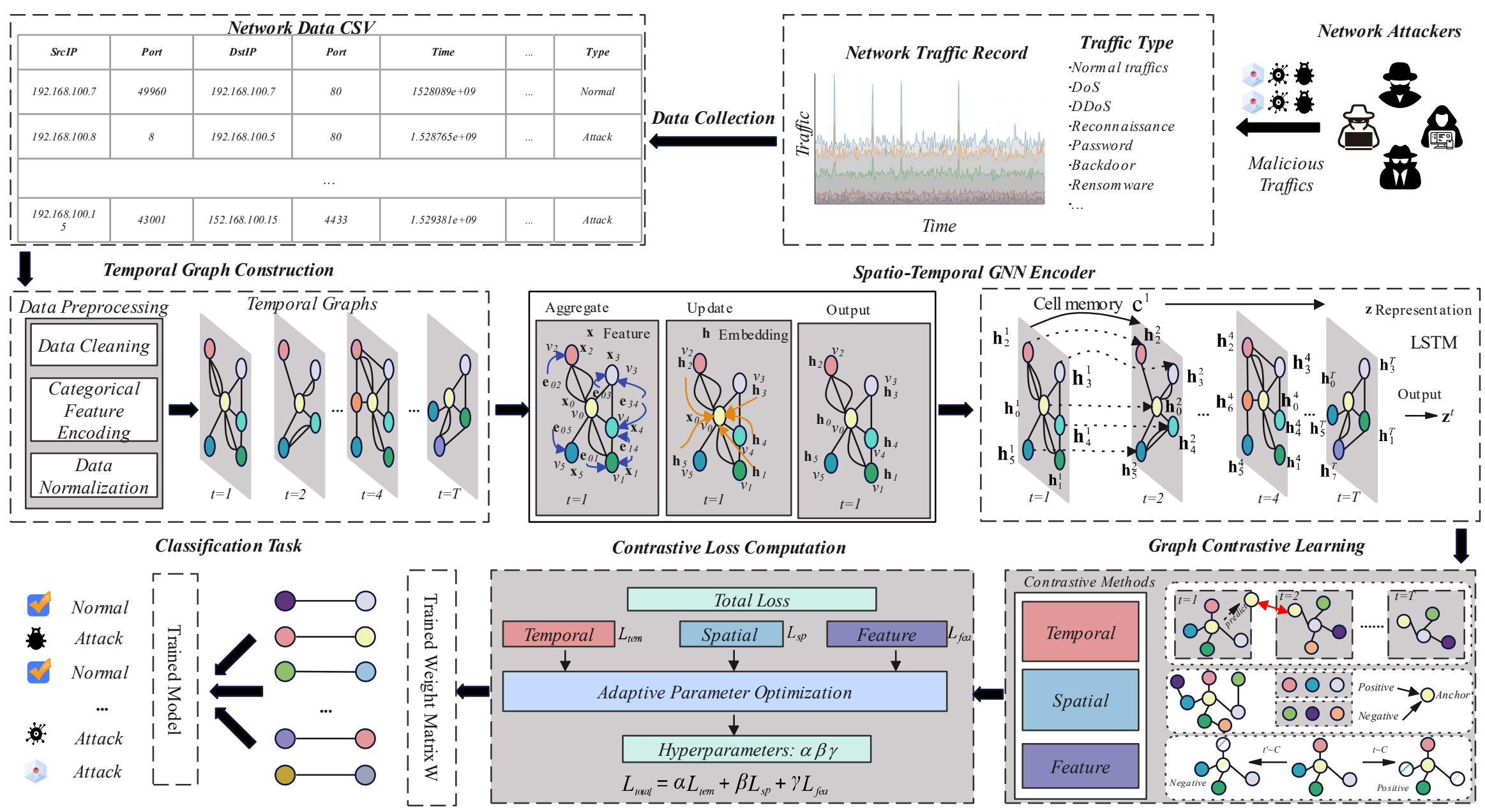

**Figure 2:** The overview of the NIDS framework. The upper layer presents the information of network traffic flows. The middle layer illustrates temporal graph construction, and shows how spatio-temporal GNN encodes node and edge features by modeling temporal dependencies. The lower layer introduces how graph contrastive learning guides the model to learn representations and detect malicious network traffic flows.

## 3.2. Feature Selection Engineering

To evaluate the discriminative ability of each feature with respect to the target variable, we train a random forest (RF) model and extract the feature-importance scores from the model. The scores measure the contribution of each feature in the process of constructing the decision tree. We then sort these features into non-increasing order by their importance scores and retain only the top-ranked features for our model training. This helps accelerate the training process. The top-15 feature importance scores of ToN-IoT dataset are presented in Fig.3.

## 3.3. Graph Construction

Given a dataset, we first sort all network traffic flows in non-decreasing order of their timestamps, and then divide them into $T$ daily intervals. For each interval $t \in \{1, \dots, T\}$, we construct a temporal graph $\mathcal{G}^t = (\mathcal{V}^t, \mathcal{E}^t)$ using the flows occurring within the interval $t$. Here, $\mathcal{V}^t$ is the node set, where each node $v_i \in \mathcal{V}^t$ corresponds to a unique network host identified by its IP address. Note that two different graphs, $\mathcal{V}^t$ and $\mathcal{V}^{t'}$, may share some nodes. $\mathcal{E}^t$ is the edge set, where each edge $e^t_{ijk} \in \mathcal{E}^t$ represents the $k$-th network traffic flow between the nodes $v_i$ and $v_j$ during the time interval $t$. For $1 \leq t \leq T$, we define the feature of each node $v_i \in \mathcal{V}^t$ as the vector $\mathbf{x}^t_i$, and the feature of each edge $e^t_{ijk} \in \mathcal{E}^t$ as the vector $\mathbf{e}^t_{ijk}$. Furthermore, given two temporal graphs $\mathcal{G}^t$ and $\mathcal{G}^{t+t_1}$, we call them adjacent if $t_1 = 1$, i.e., $\mathcal{G}^{t+t_1}$ is in the next time interval. In contrast, we call them distant if $t_1$ is sufficiently large. In our experiments, $t_1$ is set to 4 for distant graphs.

## 3.4. Spatio-Temporal GNN Encoder

After graph construction for the dataset, we obtain a sequence of temporal graphs $\{\mathcal{G}^t\}_{t=1}^{T}$. We now propose a novel spatio-temporal GNN encoder to learn the representation $\mathbf{z}^t_i$ of each node $v_i \in \mathcal{V}^t$ of every graph $\mathcal{G}^t$, for $1 \leq t \leq T$.

The main idea of the proposed encoder is that: the representation of each node of temporal graph $\mathcal{G}^t$ is obtained using the standard LSTM framework on the node and edge features of $\mathcal{G}^t$ and the representations of nodes in the previous temporal graph $\mathcal{G}^{t-1}$. The design enables the model to capture the temporal dependencies of network flows across consecutive timestamp-ordered graphs. For the temporal graph $\mathcal{G}^1$, the hidden state $\mathbf{h}_i^0$ and cell memory $\mathbf{c}_i^0$ for each node $v_i$ are initialized to zero due to the absence of historical information. For a subsequent temporal graph $\mathcal{G}^t$, $2 \leq t \leq T$, we assume that the node representations for all previous graphs $\mathcal{G}^{t'}$ with $t' < t$ have already been obtained. The node representations for $\mathcal{G}^t$ are computed in the following steps. For every node $v_i \in \mathcal{V}$, the encoder first concatenates the initial node features $\mathbf{x}_i^t$ with the previous embedding $\mathbf{h}_i^{t-1}$, and then maps the concatenated vector to the same dimension space as the node embedding $\mathbf{h}_i^t$ through a learnable weight matrix $W$ followed by a nonlinear activation function $\sigma$. That is:

$$\mathbf{h}_i^t = \sigma(W \cdot [\mathbf{x}_i^t, \mathbf{h}_i^{t-1}]) \tag{1}$$

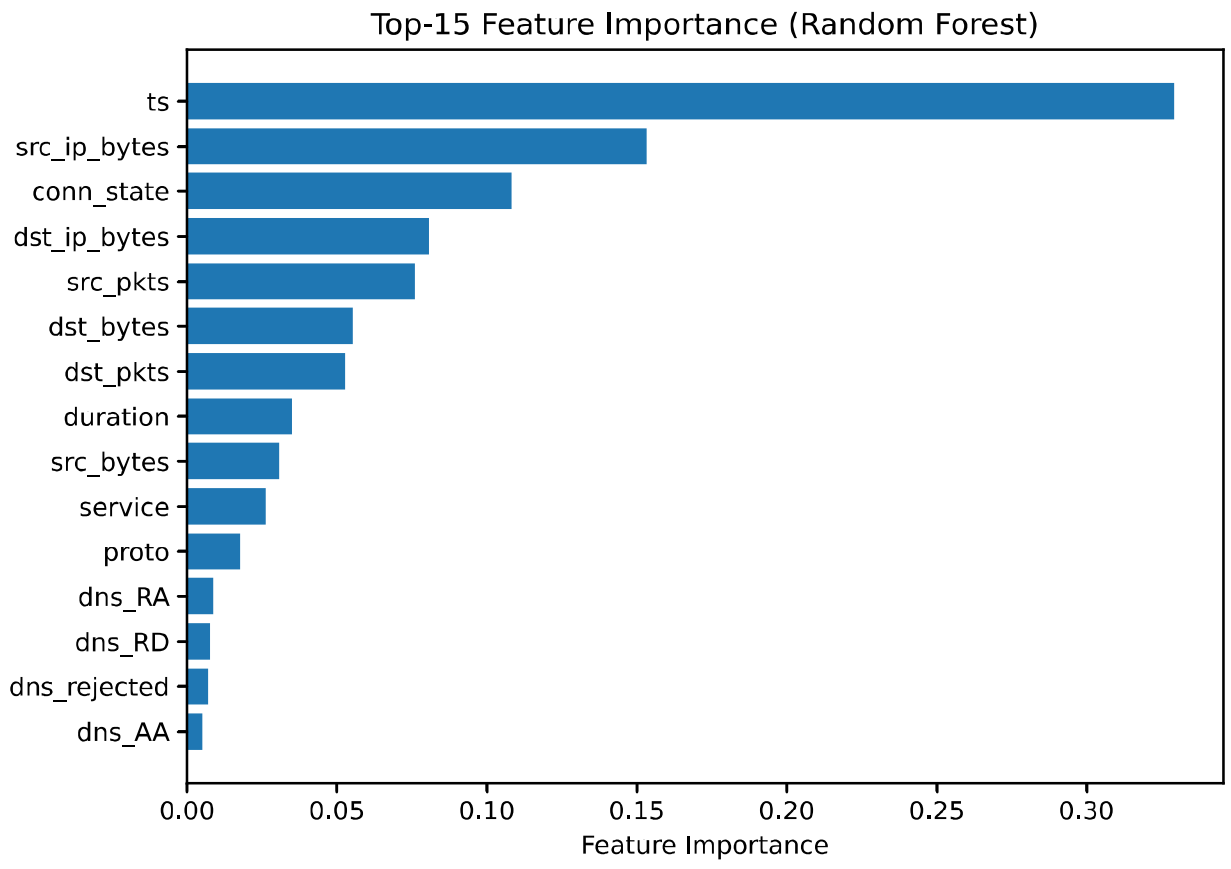

**Figure 3:** The Top-15 feature importance of ToN-IoT Dataset.

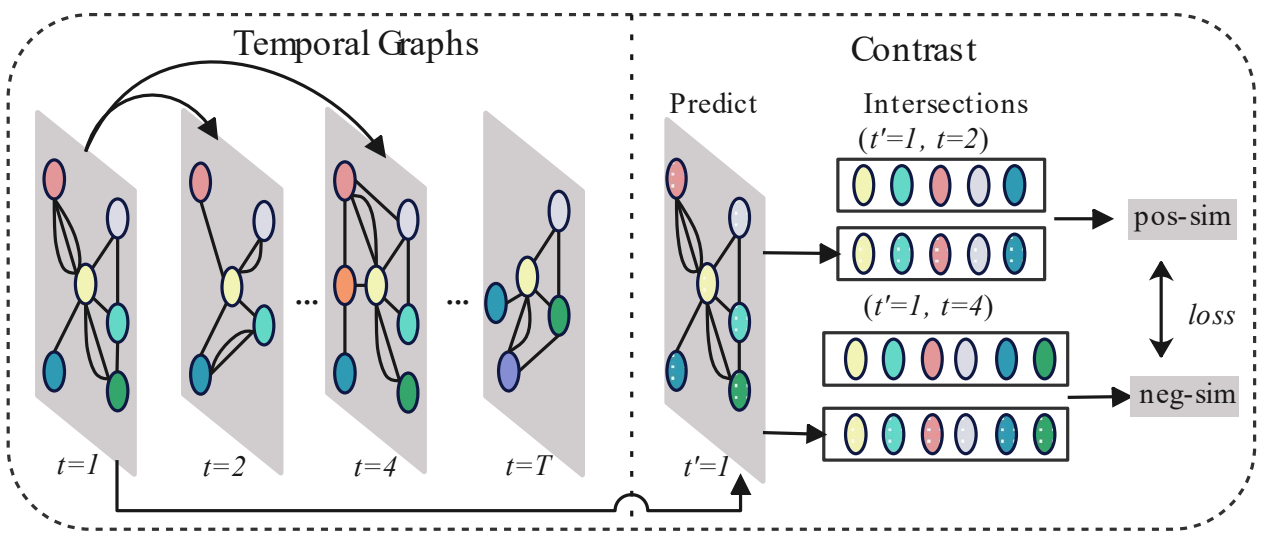

**Figure 4:** The temporal contrastive learning process between nodes across different temporal graphs.

, where $\mathbf{h}_i^t$ will be used as the input of E-GraphSAGE. This design enables the model to simultaneously consider the inherent features and historical state information of nodes, providing a richer input representation for subsequent graph encoding operations.

Subsequently, in E-GraphSAGE with $l$-layer aggregation, the initial node embedding $\mathbf{h}_i^{t,0}$ of each node $v_i \in \mathcal{V}$ is set to $\mathbf{h}_i^t$. For $l > 0$, the node embedding $\mathbf{h}_i^{t,l}$ of each node $v_i$ in time interval $t$ can be obtained as Equation (2):

$$\mathbf{h}_i^{t,l} = \sigma(W_l \cdot \mathrm{CONCAT}(\mathbf{h}_i^{t,l-1}, \mathbf{h}_{N(i)}^{t,l})). \tag{2}$$

Here, $W_l$ is the weight matrix, $\sigma$ is the activation function, and $\mathbf{h}_{N(i)}^{t,l}$ is the aggregated representation of neighbors of $v_i$. It is computed as shown in Equation (3):

$$\mathbf{h}_{N(i)}^{t,l} = \mathrm{AGG}_l(\mathbf{h}_j^{t,l-1}, \mathbf{e}_{ijk}^{t,l-1}) \tag{3}$$

, where $\mathrm{AGG}_l$ is the aggregation operation of each layer $l$, , which is set to mean aggregation in our implementation. Here, $\mathbf{h}_j^{t,l-1}$ is the node embeddings $\mathbf{h}_j^t$ in layer $l-1$ and $\mathbf{e}_{ijk}^{t,l-1}$ is the edge features $\mathbf{e}_{ijk}^t$ of the $k$-th edge between nodes $v_i$ and $v_j$ in layer $l-1$ within time interval $t$. The final node embedding $\mathbf{h}_i^t$ for each node $v_i$, i.e., the output of E-GraphSAGE module, can be obtained through a softmax layer.

Finally, LSTM regulates information flow via gating mechanism, which helps alleviate the vanishing gradient problem in recurrent sequence modeling and enable the learning of long-range temporal dependencies. With the input of E-GraphSAGE module, the node embeddings $\mathbf{h}_i^t$ and cell memory $\mathbf{c}_i^t$ of LSTM module can be expressed as Equation (4).

$$(\mathbf{h}_i^t, \mathbf{c}_i^t) = \mathrm{LSTM}(\mathbf{h}_i^t, (\mathbf{h}_i^{t-1}, \mathbf{c}_i^{t-1})) \tag{4}$$

, where $\mathbf{c}_i^{t-1}$ is the cell memory of each node $v_i$ in time interval $t-1$. After encoding, we can obtain the final node representation $\mathbf{z}_i^t = \mathbf{h}_i^t$ of each node $v_i$ in time interval $t$ through a softmax layer.

## 3.5. Graph Contrastive Learning

The selection of positive and negative samples in graph contrastive learning is paramount for the classification task. Motivated by the importance of temporal information, spatial dependencies and features of nodes, three kinds of contrastive strategies are designed in our method. Through self-supervised contrastive learning strategies, the learning of node representations is constrained from these multiple views, thereby capturing temporal smoothness, spatial consistency, and feature robustness.

### *3.5.1. Temporal Contrastive Learning*

In networks, host communications often persist across consecutive time intervals, while network workloads and attack behaviors may evolve over longer time intervals [35, 36]. Thus, node states in our constructed temporal graphs tend to evolve smoothly between adjacent graphs but differ substantially across distant graphs. Motivated by this, the temporal contrastive learning loss aims to model the evolution of node representations in the temporal dimension. The whole process is presented in Fig. 4. By distinguishing the similarity of the same node within adjacent time intervals from that in distant time intervals, the model is encouraged to learn temporal smoothness. Positive samples are defined as $(v_i^t, v_i^{t+1})$ from two adjacent graphs, while negative samples are $(v_i^t, v_i^{t+4})$ from two distant graphs. To explicitly model the node evolution, we introduce a temporal predictor to learn the state transition from time interval $t$ to $t+1$, and use the predicted representation to match the representation at interval $t+1$. Specifically, given $\mathbf{z}_i^t$ for node $v_i$ in $\mathcal{G}^t$, the predictor outputs $\hat{\mathbf{z}}_i^t$ via:

$$\hat{\mathbf{z}}_i^t = \sigma(W_p \cdot \mathbf{z}_i^t) \tag{5}$$

, where $\sigma$ is ReLU and $W_p$ is a learnable weight matrix.

Meanwhile, network topologies evolve over time as nodes may appear or disappear, so the node sets of temporal graphs from different time intervals are not fully aligned. In temporal graph contrastive learning, directly contrasting all node representations may introduce noise and unnecessary overhead, since nodes without semantic correspondence cannot form effective positive or negative constraints. Therefore, we select the intersections $\mathcal{P} = \mathcal{V}^t \cap \mathcal{V}^{t+1}$, and $\mathcal{Q} = \mathcal{V}^t \cap \mathcal{V}^{t+4}$, and use the representations of these corresponding nodes to calculate the contrastive loss, which maintains semantic consistency and improves efficiency.

To model temporal consistency and discriminability of node representations, we design a joint objective consisting of a temporal alignment loss and a temporal interval constraint. Specifically, we encourage smooth and continuous evolution by minimizing an alignment loss between the predicted representation $\hat{\mathbf{z}}_i^t$ and the representation $\mathbf{z}_i^{t+1}$ of the corresponding node at the next time interval $t+1$, which is defined as Equation (6):

$$\mathcal{L}_{\text{align}} = \frac{1}{|\mathcal{P}|}\sum_{i\in\mathcal{P}}(2 - 2\hat{\mathbf{z}}_i^t \cdot \mathbf{z}_i^{t+1}). \tag{6}$$

To avoid representation collapse across different temporal graphs, we further introduce an interval separation constraint on the same nodes that are temporally distant. By enforcing an upper bound on their similarity, we maintain sufficient discriminability by Equation (7):

$$\mathcal{L}_{\text{uniform}} = \frac{1}{|\mathcal{Q}|}\sum_{i\in\mathcal{Q}}\max(0, \hat{\mathbf{z}}_i^t \cdot \mathbf{z}_i^{t+4} - y) \tag{7}$$

, where $y$ is a similarity upper bound.

The final temporal loss $\mathcal{L}_{\text{tem}}$ is:

$$\mathcal{L}_{\text{tem}} = \mathcal{L}_{\text{align}} + \psi \cdot \mathcal{L}_{\text{uniform}} \tag{8}$$

, where $\psi$ is a penalty weight that balances the alignment and separation terms.

### 3.5.2. *Spatial Contrastive Learning*

In networks, hosts that directly and frequently communicate often exist in the same subnet or service chain, and thus exhibit similar interaction patterns [9, 37]. Accordingly, the spatial contrastive learning loss leverages graph topology by constructing contrastive pairs within the local neighborhood: connected node pairs are treated as positive samples, while randomly sampled unconnected node pairs serve as negative samples. By pulling neighbors closer and pushing non-neighbors apart, the model learns node representations that capture spatial dependencies and local structural consistency.

Specifically, we take $(\mathbf{z}_i^t, \mathbf{z}_j^t)$ as a positive sample pair if there exists an edge $e_{ijk}^t \in \mathcal{E}^t$ between $v_i$ and $v_j$. For negative sampling, we use the adjacency matrix to ensure that each negative node is not directly connected to the anchor nodes $v_i$, avoiding sampling observed neighbors as negatives. For each positive sample pair, we sample $S$ negative nodes to increase the difficulty and improve the informativeness of spatial contrastive learning. The spatial contrastive loss is defined as:

$$\mathcal{L}_{\text{spa}} = -\frac{1}{|\mathcal{E}^t|}\sum_{(v_i,v_j)\in\mathcal{E}^t}\log\frac{\exp((\mathbf{z}_i^t)^\top \mathbf{z}_j^t/\tau)}{D_{ij}} \tag{9}$$

, where

$$D_{ij} = \exp((\mathbf{z}_i^t)^\top \mathbf{z}_j^t/\tau) + \sum_{s=1}^{S}\exp((\mathbf{z}_i^t)^\top \mathbf{z}_{i,s}^t/\tau).$$

Here, $\tau$ is the temperature coefficient and $\mathbf{z}_{i,s}^t$ denotes the representation of the $s$-th negative node sampled for $v_i$.

### 3.5.3. *Feature Contrastive Learning*

In practice, network flows may be collected incompletely, which can introduce noise and missing values into edge features, thereby affecting the stability of the learned representations [38]. A common strategy in contrastive learning is to create semantically consistent yet perturbed views via feature augmentation, so that the encoder learns representations that generalize better. Following this idea, we design an adaptive feature corruption strategy inspired by GCA, which injects stronger perturbations into less informative node representations while preserving semantic information. Specifically, given node representations $\mathbf{z}_i^t$, we obtain a node-wise corruption score from global structural statistics. Let $\hat{w}_i$ denote log-degree-based score averaged over time, where low-degree nodes yield larger $\hat{w}_i$ and thus receive stronger corruption. In negative perturbation, we normalize $\hat{w}_i$ by min-max scaling:

$$\hat{w}_i = \frac{w_i - w_{min}}{w_{max} - w_{min}}. \tag{10}$$

The corruption strength is then determined as:

$$\delta_i = \text{clip}(\delta_0 \hat{w}_i, \delta_{min}, \delta_{max}) \tag{11}$$

, where $\text{clip}(\delta_0 \hat{w}_i, \delta_{min}, \delta_{max}) = \min(\max(\delta_0 \hat{w}_i, \delta_{min}), \delta_{max})$ with $\delta_0 = 0.2$, $\delta_{min} = 0.01$ and $\delta_{max} = 0.3$. We generate the augmented view by interpolating the node representations with Gaussian noise:

$$\hat{\mathbf{z}}_i^+ = (1-\delta_i)\mathbf{z}_i^t + \delta_i \epsilon_i^+, \epsilon_i^+ \mathcal{N}(0, \eta_+^2 \mathbf{I}) \tag{12}$$

, where $\eta_+ = 0.1$. This design applies stronger perturbations to nodes with larger corruption scores.

In negative perturbation, we impose stronger noise perturbations on high-degree nodes:

$$\hat{\mathbf{z}}_i^- = (1-\xi_i)\mathbf{z}_i^t + \xi_i \epsilon_i^-, \epsilon_i^- \mathcal{N}(0, \eta_-^2 \mathbf{I}) \tag{13}$$

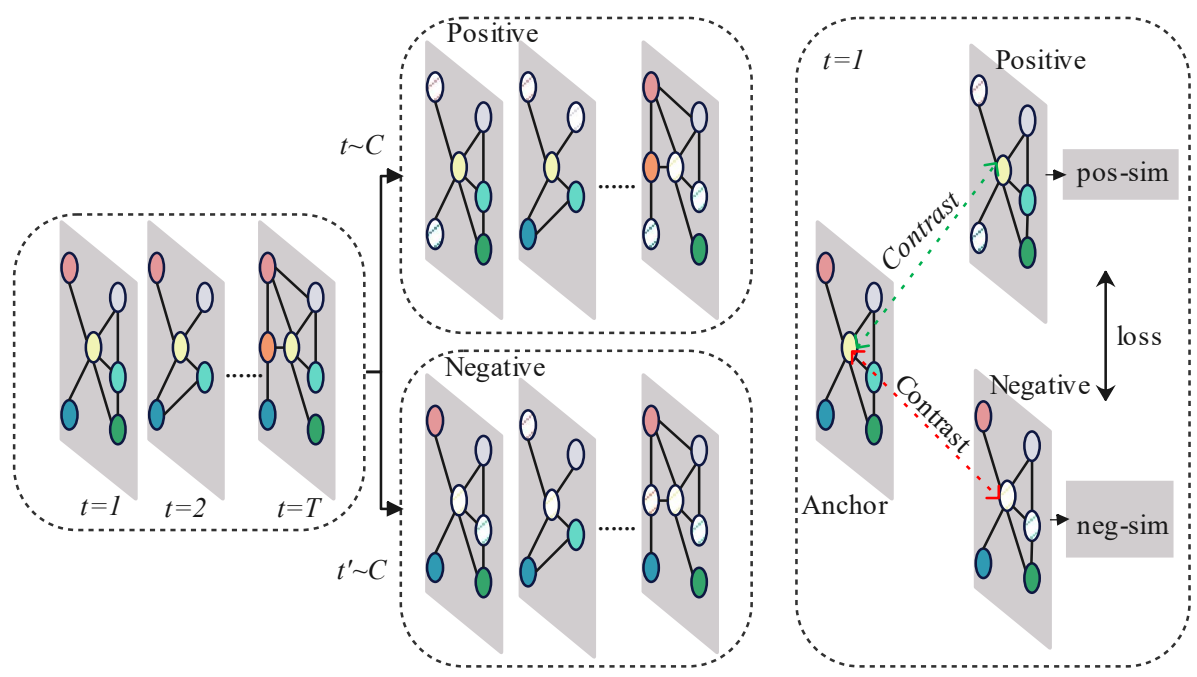

**Figure 5:** The feature contrastive learning process between nodes in the positive and negative graphs.

, where $\eta_- = 0.3$ and $\xi_i = \text{clip}(\xi_0 \cdot \frac{w_{max}-w_i}{w_{max}-w_{min}}, \xi_{min}, \xi_{max})$, with $\xi_0 = 0.5$ and $\xi_{min} = 0.05$, and $\xi_{max} = 0.5$.

The augmented positive and negative samples are utilized to compute the feature contrastive loss through normalized cosine similarity and cross-entropy:

$$\mathcal{L}_{\text{fea}} = -\frac{1}{N} \sum_{i=1}^{N} \log A_i \tag{14}$$

, where

$$A_i = \frac{\exp((\mathbf{z}_i^t, \hat{\mathbf{z}}_i^+)/\tau)}{\exp((\mathbf{z}_i^t, \hat{\mathbf{z}}_i^+)/\tau) + \exp(\text{sim}(\mathbf{z}_i^t, \hat{\mathbf{z}}_i^-)/\tau)}.$$

Here, $\tau$ is the temperature coefficient. The process of feature contrastive loss can be seen in Fig. 5. This enables the learned representation to be robust to feature missing and noise, which can enhance the distinguisability of different node representations.

### 3.5.4. Adaptive Parameter Optimization

In multi-view self-supervised learning framework, different loss usually impose distinct semantic constraints and exhibit significantly different optimization scales. Using fixed weights to combine multiple self-supervised objectives often leads to gradient domination, where one loss overwhelms the parameter updates and suppresses other learning signals. Thus, we adopt a gradient-norm based adaptive loss weighting strategy. Specifically, $\Theta_s$ denotes the shared parameter set used for learning unified representations, and $\mathcal{L}_{\text{tem}}, \mathcal{L}_{\text{spa}}$ and $\mathcal{L}_{\text{fea}}$ denote the three different losses, respectively. At each training iteration, the gradient of each loss with respect to $\Theta_s$ is computed and its $\ell_2$ norm is used to measure the optimization strength of the corresponding self-supervised signal. For a loss $\mathcal{L}$, the gradient norm is defined as:

$$g = \left\| \nabla_{\Theta_s} \mathcal{L} \right\|_2 = \sqrt{\sum_{\theta \in \Theta_s} \left\| \frac{\partial \mathcal{L}}{\partial \theta} \right\|_2^2}. \tag{15}$$

Consequently, $g_{\text{tem}}$, $g_{\text{spa}}$, and $g_{\text{fea}}$ denote the gradient norms of $\mathcal{L}_{\text{tem}}$, $\mathcal{L}_{\text{spa}}$, and $\mathcal{L}_{\text{fea}}$, respectively. Accordingly, the balancing coefficients $\alpha$, $\beta$, $\gamma$ are computed dynamically based

**Algorithm 1**

**Input:** A graph set $\mathcal{G} = \{\mathcal{G}^1, \mathcal{G}^2, \ldots, \mathcal{G}^T\}$ and epochs $E$.
For $1 \le t \le T$, $\mathcal{G}^t = (\mathcal{V}^t, \mathcal{E}^t)$ is a temporal graph.
**Output:** Trained parameters $\Theta$.
1: **for** $t = 1$ to $T$ **do**
2: Obtain $z^t$ for each graph $\mathcal{G}^t$ by Eqs. (2) and (4).
3: **end for**
4: **for** $t = 1$ to $T$ **do**
5: Compute predicted node representations by Eq.(5).
6: Select intersection $\mathcal{P}$ of $\mathcal{G}^t$ and $\mathcal{G}^{t+1}$.
7: Select intersection $\mathcal{Q}$ of $\mathcal{G}^t$ and $\mathcal{G}^{t+4}$.
8: Compute similarities of $(v_i^t, v_i^{t+1})$ in $\mathcal{P}$ by Eq.(6).
9: Compute similarities of $(v_i^t, v_i^{t+4})$ in $\mathcal{Q}$ by Eq.(7).
10: Compute temporal contrastive loss $\mathcal{L}_{\text{tem}}$ by Eq. (8).
11: **for all** nodes $v_i \in \mathcal{V}^t$ **do**
12: Select neighbors as positive $z_j^t$ and non-neighbors as negative $z_{i,s}^t$.
13: Compute spatial contrastive loss $\mathcal{L}_{\text{spa}}$ by Eq. (9).
14: **end for**
15: **for all** nodes $v_i \in \mathcal{V}^t$ **do**
16: Obtain positive $\hat{z}_i^+$ and negative $\hat{z}_i^-$.
17: Compute feature contrastive loss $\mathcal{L}_{\text{fea}}$ by Eq. (14).
18: **end for**
19: Adaptive parameter optimization to obtain $\alpha, \beta, \gamma$ by Eq. (15).
20: $\mathcal{L} = \alpha\mathcal{L}_{\text{tem}} + \beta\mathcal{L}_{\text{spa}} + \gamma\mathcal{L}_{\text{fea}}$.
21: Update model parameters $\Theta$.
22: **end for**

on $g_{\text{tem}}$, $g_{\text{spa}}$, and $g_{\text{fea}}$:

$$(\alpha, \beta, \gamma) = \frac{g_{tem}, g_{spa}, g_{fea}}{g_{tem} + g_{spa} + g_{fea}} \tag{16}$$

The overall training objective is formulated as:

$$\mathcal{L}_{\text{total}} = \alpha\,\mathcal{L}_{\text{tem}} + \beta\,\mathcal{L}_{\text{spa}} + \gamma\,\mathcal{L}_{\text{fea}}. \tag{17}$$

This gradient-norm based weighting mechanism plays a crucial role in self-supervised learning. It explicitly reflects the optimization scale of each contrastive objective on the shared representation space, preventing any single loss from dominating the training process.

The whole training process is shown in Algorithm 1.

## 3.6. Classification Task

The proposed method aims to differentiate between normal and malicious behaviors and identify the specific attack classes in a self-supervised way, which can be seen as classification tasks. All datasets divided into training and testing data are the inputs of the encoders to obtain final embeddings, respectively, for further classification tasks. Widely used for the binary and multiclass classifications in NIDS, two functions, e.g., binary cross-entropy (BCE) function and the cross-entropy (CE) function, are utilized in our method to implement the classification experiments, respectively. The detailed parameters are demonstrated in Table 1.

## 3.7. The Time Complexity Analysis

The running time of our model is determined by its two modules: the spatio-temporal encoder and graph contrastive learning.

**Table 1**
Hyperparameter

| Hyperparameter | Values |
|---|---|
| Number of temporal graph $T$ | 10 |
| Learning rate (lr) | 1e-3 |
| Weight decay | 3e-5 |
| Activation function $\sigma$ | ReLU |
| Optimizer | Adam |
| Similarity upper bound $\gamma$ | 0.1 |
| Penalty weight $\psi$ | 0.05 |
| Number of negative edge $S$ | 5 |
| Temperature coefficient $\tau$ | 0.5 |

First, we consider the time complexity of spatio-temporal encoder. In the aggregation process of E-GraphSAGE, each node needs to consider its adjacent edges along with their nodes, and thus each edge is counted twice. Since $T$ temporal graphs are constructed and the edge sets of different temporal graphs are disjoint, the total number of edges across all temporal graphs is $m$, e.g., $\sum_{t=1}^{T} |\mathcal{E}^t| = m$. Hence, the aggregation process takes $O(2m)$ time. In the embedding process of LSTM, each node passes through four gates, and each gate involves a matrix multiplication whose running time depends on the input dimension $d$ of node feature. That is, the embedding process costs $O(4dn)$ time, where $n$ is the number of node in the graph. Consequently, the time complexity of the spatio-temporal encoder is $O(2m) + T \cdot O(4dn) = O(2m + 4Tdn)$, where $T$ is the number of the constructed temporal graphs.

Then, in graph contrastive learning, we design three contrastive schemes. In the temporal contrastive learning, we first compute the predicted node representations via a linear layer in $O(dn)$. We then find the node intersections among the predicted graph, the adjacent temporal graph, and the distant temporal graph in $O(3n)$. For each node in the intersections, we finally compute two similarities between positive samples and negative samples using the dot production in $O(2dn)$. Hence, the total time complexity of temporal contrastive learning is $O(dn + 3n + 2dn) = O(3n + 3dn)$. In the spatial contrastive learning, for each node in the current temporal graph, we compute the mean dot-product similarity to its neighbor nodes (positive sample) and to its $S$ ($S$ is a constant) randomly selected non-neighbor nodes (negative sample), respectively. The similarity computation consumes time $O(dn+Sd)$. In feature contrastive learning, we generate an augmented (positive) and an perturbed (negative) representation for each node via two clip-based transformations, whose temporal weights are precomputed during temporal graph construction. Therefore, the clip transformation costs $O(2n)$. We then compute the dot-product similarity between two representations for each node in $O(dn)$. Overall, the time complexity of feature contrastive learning is $O(2n + dn)$.

Since every temporal graph requires computing all these three contrastive losses, the overall cost of graph contrastive learning is $T \cdot (O(3dn + 3n) + O(dn + Sd) + O(2n + dn)) = O(5Tdn + 5Tn + TSd)$. Therefore, the time complexity of the proposed model is $O(2m + 9Tdn + 5Tn + TSd)$.

**Table 2**
Comparison of our method and other GNN-based methods on time complexity.

| Method | Time complexity |
|---|---|
| Ours | $O(2m + 9Tdn + 5Tn + TSd)$ |
| NEGAT + NEGSC | $O(18m + 2n)$ |
| TCG-IDS | $O(16m + Tn + 5Tnd + Td\bar{d}n)$ |

NEGAT + NEGSC has a time complexity of $O(4KIm + 2Im + 2n)$, where $O(4KIm + 2Im)$ is by the computation of attention coefficients and aggregation of NEGAT, and $I$ is the numbers of attention heads and $K$ is the number of layers, and $O(2n)$ is by the edge construction step of NEGSC. With heads $I = 3$ and layer $K = 1$, the time complexity is $O(18m+2n)$. TCG-IDS has a time complexity of $O(2KIm + Tn + Tdn + Td\bar{d}n + 4Tdn)$, where $O(4KIm + Tn)$ is by the temporal encoder which utilizes attention mechanism, and $O(Tdn)$, $O(Td\bar{d})$, $O(4Tdn)$ are by three contrasting strategies, and $\bar{d}$ is the dimension of predicted embeddings. With heads $I = 8$ and layer $K = 1$,, the time complexity of TCG-IDS is $O(16m + Tn + 5Tdn + Td\bar{d}n)$.

In the constructed temporal graphs of NIDS, nodes represent hosts, and edges represent communications between them. In networks, the number of edges is typically much larger than that of the nodes, as hosts may communicate with each other frequently. Thus, the temporal graphs are often dense, leading to $m \gg n$, i.e., $m$ may grow on $O(n^2)$ or even faster. Consequently, the overall time complexity is dominated by terms involving $m$, while the node-wise terms (e.g., $O(n)$ or $O(dn)$) are comparatively minor. In this setting, our method runs in $O(2m)$, which achieves the lowest time complexity among the related representative methods. This advantage arises from explicitly modeling temporal dependencies, which avoids the edge-dependent computational overhead introduced by attention mechanisms in other methods. The details are listed in Table 2.

# 4. Experiments

In this section, we first present the experimental setup, datasets and metrics. The performance of the proposed model is then reported for binary and multiclass classifications. Moreover, comparisons with existing methods are conducted to evaluate the effectiveness of our approach. Finally, ablation studies are provided to demonstrate the importance of temporal information.

## 4.1. Setup

Our model is implemented on a single NVIDIA GeForce RTX 4090 with 24GB of GPU memory using Python, PyTorch [39] and PyTorch Geometric [40]. We split the datasets using a 7:3 ratio for training and testing. After preprocessing the raw data, 70% of the data for training is processed and fed into the model, which enables the model to learn effective information through iteratively optimizing and tuning. The remaining 30% of the data for testing is utilized for evaluating the performance and efficiency of the model.

**Table 3**
Datasets Statistics.

| Dataset | Normal Flows | Attack Flows | Classes | Flow Features | Time Attribute |
|---|---|---|---|---|---|
| BoT-IoT | 477 | 3,668,045 | 5 | 46 | stime |
| ToN-IoT | 788,599 | 21,190,031 | 10 | 45 | ts |
| UNSW-NB15 | 2,218,764 | 321,283 | 10 | 49 | stime |
| NF-UNSW-NB15-v3 | 2,237,731 | 127,693 | 10 | 49 | FLOW START MILLISECONDS |

## 4.2. Datasets

In our study, three public and widely used datasets are chosen to implement the experiments, which are BoT-IoT [41], ToN-IoT [42], and UNSW-NB15 [43]. Recently, netflow datasets that incorporate the temporal features are emerged to address the gap in standardised feature sets, e.g., NF-BoT-IoT-v3, NF-ToN-IoT-v3, NF-UNSW-NB15-v3 [44], etc. For evaluating the efficiency of our method in the netFlow datasets, we select NF-UNSW-NB15-v3 as a distinctive example. The time-relevant features (e.g., timestamp, durations) in the raw data are crucial for capturing the dynamic nature of network traffic in the real-world scenarios. BoT-IoT, ToN-IoT, UNSW-NB15 and NF-UNSW-NB15-v3 are all the datasets used for intrusion detection. The details of four datasets are demonstrated on the Table 3, which includes the labels, explicit attack types and its corresponding amounts of each type contained in the raw data. The four datasets are all fit for binary classification, which is to distinguish whether the traffic is anomalous, and multiclass classification, which is to identify the certain type of the attacks.

The BoT-IoT dataset is created by designing a realistic network environment in the Cyber Range Lab of UNSW Canberra, which incorporated a combination of normal and botnet traffic. It consists of 46 basic features especially start time and late time of each traffic, and the amount of network traffics are 3,668,522, out of which 447 are normal, and 3,668,045 are attacks which predominantly includes 4 specific attack types, e.g., DDoS, DoS, and Reconnaissance, and another attack theft which is a small percentage at all samples. Attack category and subcategory attributes are introduced in the dataset, which could be used to train and test the model in both binary and multiclass classifications.

The ToN-IoT dataset is collected from a realistic and large-scale network designed at the Cyber Range and IoT Labs, which can be utilized for evaluating the fidelity and efficiency of different cybersecurity applications such as intrusion detection systems based on artificial intelligence. It contains 45 basic features including timestamp (ts), and the amount of network traffic are 21,978,630, out of which 788,599 are normal, and 21,190,031 are attacks, the specific types of which are Scanning, DDoS, DoS, Xss, Password, Backdoor, Injection, Ransomware and Mitm.

The UNSW-NB15 is a dataset developed at UNSW Canberra, which is designed based on a synthetic environment for generating attack activities. It is made up of 49 basic features especially timestamp of each traffic, and the amount of network traffics are 2,540,047, out of which 2,218,764 are normal, and 321,283 are attacks which includes 9 uneven specific attack types, e.g., Generic, Exploits, Fuzzers, DoS, Reconnaissance, Analysis, Backdoor, Shellcode and Worms. The purpose of the produced dataset is to be utilized for the generation and validation of intrusion detection.

The NF-UNSW-NB15-v3 dataset is the third iteration of NetFlow-based datasets converted from the UNSW-NB15 [43], which added time-related features like FLOW START MILLISECONDS, FLOW START MILLISECONDS, and SRC TO DST IAT MIN, etc. The inclusion of timestamp information allows for identifying the exact time of the traffic whne the original traffic was captured [44], which poses a new challenge to devle into temporal analysis of datasets in NIDS. Specifically, it contains 2,365,424 flows in total, out of which 2,365,424 are benign, and 127,693 are malicious flows, which has the same specific attack types as the UNSW-NB15. Deeper insights of dynamic patterns and temporal features of network behavior can be explored in the netflow dataset due to the added temporal features.

## 4.3. Evaluation Metrics

To evaluate the performance of our model in binary and multiclass classifications, five metrics widely used in other recent studies are selected, which are $Accuracy$, $Precision$, $F1-Score$, and $Recall$. The calculation of all the evaluation metrics is based on four factors, namely, True Positive ($TP$), a case where the model correctly predicts the positive class, False Positive ($FP$), a case where the model incorrectly predicts the positive class, True Negative ($TN$), a case where the model correctly predicts the negative class, and False Negative ($FN$), a case where the model incorrectly predicts the negative class.

$Accuracy$ is the rate at which the model correctly predicts in all samples.

$$Accuracy = \frac{TP + TN}{TP + FP + TN + FN} \tag{18}$$

$Precision$ measures the proportion of the samples predicted by the model to be attacks are actual attacks.

$$Precision = \frac{TP}{TP + FP} \tag{19}$$

$Recall$ represents the ratio of positive samples correctly predicted by the model to all samples that are actual positive samples.

$$Recall = \frac{TP}{TP + FN} \tag{20}$$

$F1 - Score$ is the harmonic mean of Precision and Recall, which is used to balance the two metrics.

$$F1 - Score = 2 \times \frac{Precision \times Recall}{Precision + Recall} \tag{21}$$

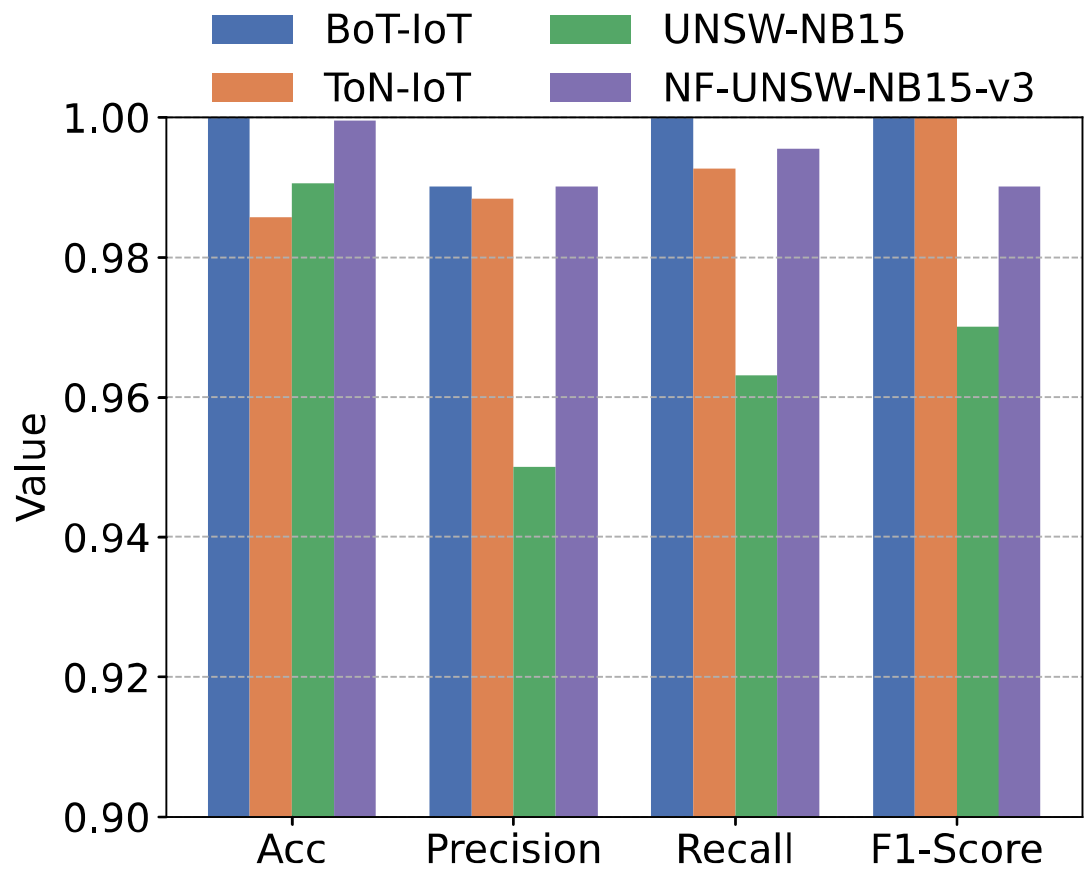

**Figure 6:** The results of binary classifications on four datasets.

## 4.4. Experiment Results

The experiment results on four datasets (e.g., BoT-IoT, ToN-IoN, UNSW-NB15, and NF-UNSW-NB15-v3) for both binary and multiclass classifications are illustrated in four separately subsection. To evaluate the efficiency and performance of our proposed method, we compare the method with existing graph-based methods, which are widely applied in NIDS. Furthermore, the ablation experiments are implemented to study the importance of spatial dependencies, temporal information and semantic features, following the corresponding results.

### 4.4.1. *Binary Classification*

The BoT-IoT, ToN-IoT, UNSW-NB15 and NF-UNSW-NB15-v3 datasets contain numerous flow data, which is more challenging for the proposed method to evaluate the detection ability between normal and malicious flows. As the same operation in [9], we use full datasets BoT-IoT, UNSW-NB15 and NF-UNSW-NB15-v3, and randomly sample 10% IoN-IoT dataset due to its large size. The overall performance of binary classifications on four datasets are illustrated in Figure. 6.

Specifically, on the BoT-IoT dataset, our method achieves 99.99% accuracy, 0.99 precision, 99.99% recall, and 0.99 F1-score. Similarly, on the ToN-IoT dataset, the model obtains 98.57% accuracy, 0.98 precision, 99.26% recall, and 0.99 F1-score. Such high performance on two IoT-related datasets can be attributed to the ability of the proposed framework to jointly model spatial dependencies among hosts and temporal information of traffic flows, enabling effective detection of attacks.

Furthermore, on the UNSW-NB15 dataset, our method achieves 99.05% accuracy, 0.95 precision, 96.31% recall, and 0,97 F1-score. On the NF-UNSW-NB15-v3 dataset, the method further reaches 99.95% accuracy and 0.99 F1-score, confirming that explicitly exploiting temporal information helps capture evolving attack behaviors.

**Table 4**
Multiclass classification statistics and results of specific classes in BoT-IoT.

| Classes | Statistics | | Metrics | |
|---|---|---|---|---|
| | Samples | Percentages | Recall | F1-Score |
| Normal | 477 | 0.00% | 80.42% | 0.86 |
| DoS | 1,620,260 | 44.16% | 99.89% | 0.99 |
| DDoS | 1,926,624 | 52.51% | 99.98% | 0.99 |
| Reconnaissance | 91,082 | 2.48% | 99.90% | 0.99 |
| Theft | 79 | 0.00% | 0.00% | 0.00 |
| **Weighted Average** | - | - | **99.93%** | **0.99** |

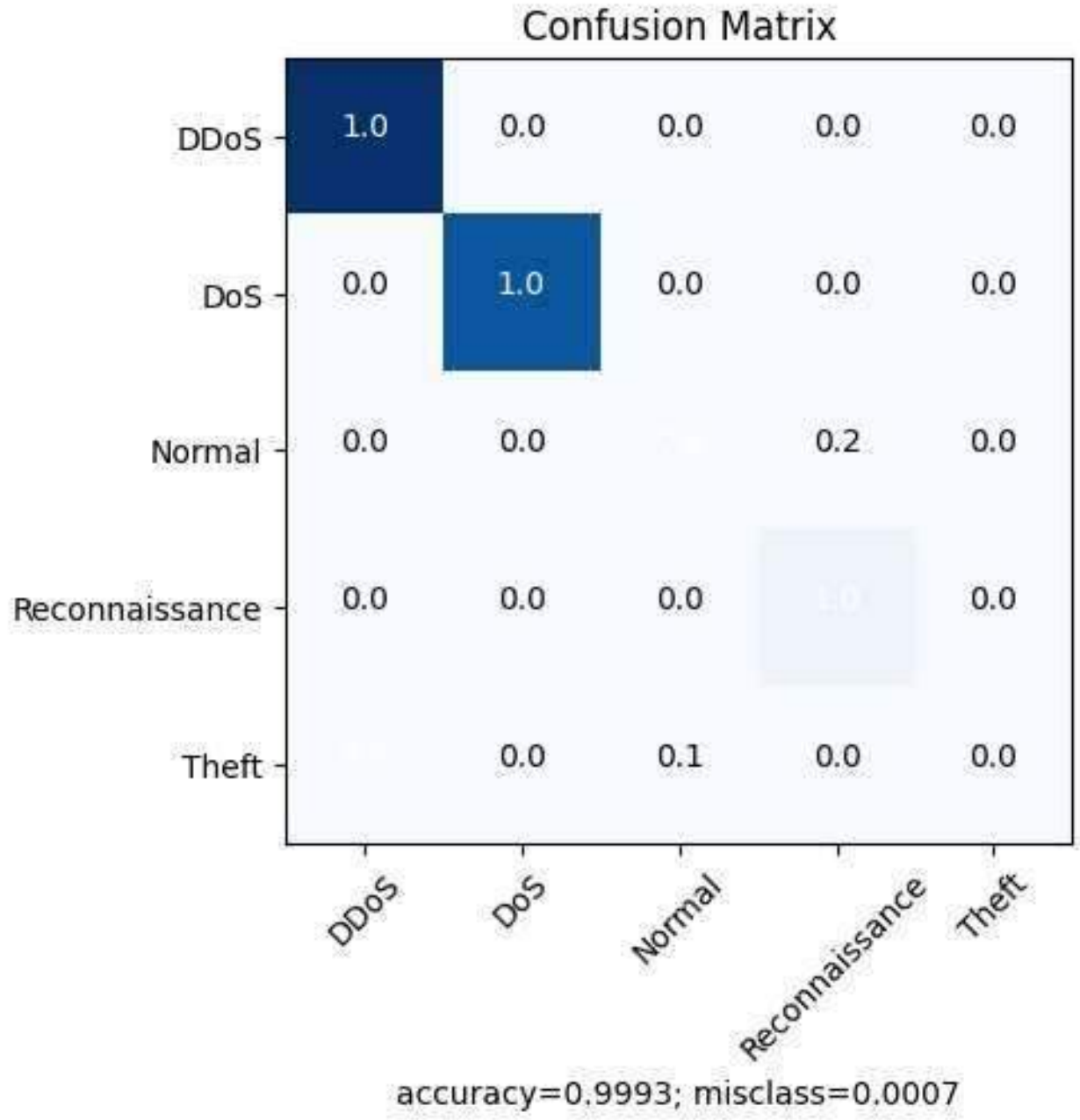

**Figure 7:** The confusion matrix of BoT-IoT in multiclass classification.

The results on four datasets indicate that integrating spatial dependencies and temporal information is crucial for reliable detection across heterogeneous network scenarios.

### 4.4.2. *Multiclass Classification*

After delving into the performance of our method on binary classifications, the performance of multiclass classifications evaluated on four datasets (e.g.,BoT-IoT, ToN-IoT, UNSW-NB15 and NF-UNSW-NB15-v3) are also worth discussing.

Specifically, on the BoT-IoT dataset, the performance evaluated on four classes are superior which has a F1-Score of 0.86, 0.99, 0.99, and 0.99 on Normal, DoS, DDoS, and Reconnaissance, respectively, and has a weighted average recall of 99.93%. The perfect performance on DoS and DDoS indicates that the model successfully captures burst temporal patterns, which are key features of flooding attacks. The only underperformance was in an attack Theft due to its few samples in the dataset, only accounting for almost 0.00% in all samples. The statistics and detailed results of specific classes in BoT-IoT are illustrated in Table 4 and Fig.7.

**Table 5**
Multiclass classification statistics and results of specific classes in ToN-IoT.

| Classes | Statistics | | Metrics | |
|---|---|---|---|---|
| | Samples | Percentages | Recall | F1-Score |
| Normal | 78,906 | 3.72% | 62.82% | 0.73 |
| DoS | 338,037 | 15.95% | 96.81% | 0.78 |
| DDoS | 616,683 | 29.10% | 92.30% | 0.95 |
| Backdoor | 50,873 | 2.40% | 0.00% | 0.00 |
| Injection | 45,150 | 2.13% | 28.86% | 0.44 |
| Password | 135,766 | 6.17% | 82.68% | 0.74 |
| Ransomware | 80,022 | 3.64% | 63.33% | 0.67 |
| Scanning | 714,561 | 32.51% | 85.62% | 0.89 |
| Xss | 210,566 | 9.58% | 87.63% | 0.83 |
| Mitm | 1,156 | 0.05% | 0.01% | 0.02 |
| **Weighted Average** | - | - | **84.52%** | **0.83** |

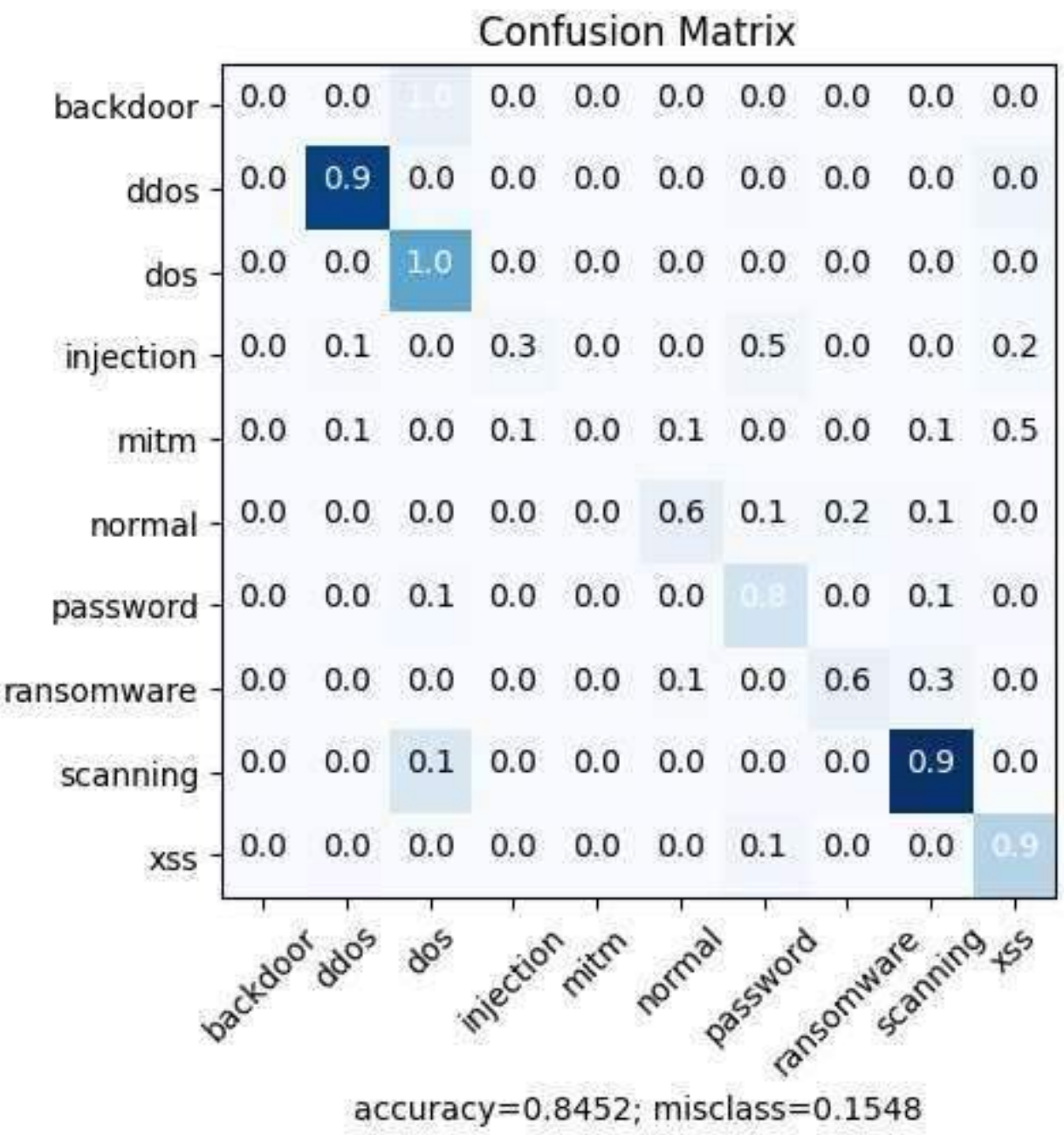

**Figure 8:** The confusion matrix of ToN-IoT in multiclass classification.

On the ToN-IoT dataset, the result of our method is comparatively general due to the its complex and unweighted classes, especially for DoS, DDoS, Scanning and Mitm attacks. Still, the performance evaluated in most classes are terrific, which has a F1-Score of 0.95, 0.89 and 0.83 in DDoS, Scanning and Xss, and a weighted average recall of 84.52%, which demonstrate the effectiveness in identifying diverse attacks in the complex network scenarios. From the confusion matrix 8, these attacks typically involve host communications and temporal behavior, which can be effectively captured through the proposed contrastive learning method. The statistics and detailed results of specific classes in ToN-IoT are illustrated in Table 5 and Fig.8.

For UNSW-NB15 and its netflow version 3 NF-UNSW-NB15-v3 datasets with the same and unweighted classes, there is a high percentage of benign flows, accounting for 87.35% and 94.60%, highly susceptible to inaccuarate results. The performance of our model on UNSW-NB15 and NF-UNSW-NB15-v3 have a weighted average recall of 98.31% and 97.81%, and F1-Score of 0.98 and 0.97, respectively The performance indicates that the model effectively leverages temporal evolution to differentiate benign fluctuations from malicious behaviors, while exploiting spatial dependencies to discriminate among diverse attack types. The statistics and detailed results of specific classes in UNSW-NB15 and NF-UNSW-NB15-v3 are illustrated in Table 6 and 7, respectively.

**Table 6**
Multiclass classification statistics and results of specific classes in UNSW-NB15.

| Classes | Statistics | | Metrics | |
|---|---|---|---|---|
| | Samples | Percentages | Samples | Percentages |
| Normal | 2,218,764 | 87.35% | 100% | 1.00 |
| Analysis | 2,677 | 0.10% | 0.00% | 0.00 |
| Backdoor | 2,329 | 0.09% | 0.00% | 0.00 |
| DoS | 16,353 | 0.64% | 0.45% | 0.00 |
| Exploits | 44,525 | 1.75% | 90.89% | 0.69 |
| Fuzzers | 24,246 | 0.95% | 76.20% | 0.76 |
| Generic | 215,481 | 8.48% | 97.42% | 0.98 |
| Reconnaissance | 13,987 | 0.55% | 66.99% | 0.68 |
| Shellcode | 1,511 | 0.05% | 0.00% | 0.00 |
| Worms | 174 | 0.00% | 0.00% | 0.00 |
| **Weighted Average** | - | - | **98.31%** | **0.98** |

**Table 7**
Multiclass classification statistics and results of specific classes in NF-UNSW-NB15-v3.

| Classes | Statistics | | Metrics | |
|---|---|---|---|---|
| | Samples | Percentages | Recall | F1-Score |
| Benign | 2,237,731 | 94.60% | 99.99% | 0.99 |
| Analysis | 1,226 | 0.05% | 0.00% | 0.00 |
| Backdoor | 4,659 | 0.19% | 0.00% | 0.00 |
| DoS | 5,980 | 0.25% | 0.56% | 0.01 |
| Exploits | 42,748 | 1.80% | 65.92% | 0.65 |
| Fuzzers | 33,816 | 1.42% | 82.62% | 0.62 |
| Generic | 19,651 | 0.83% | 67.09% | 0.72 |
| Reconnaissance | 17,074 | 0.72% | 39.57% | 0.47 |
| Shellcode | 2,381 | 0.10% | 0.00% | 0.00 |
| Worms | 158 | 0.00% | 0.00% | 0.00 |
| **Weighted Average** | - | - | **97.81%** | **0.97** |

### *4.4.3. Compared with Existing Graph-based Methods*

In order to intuitively understand the effectiveness of our model, we present a more comprehensive comparison with the existing graph-based methods, which includes supervised and self-supervised GNN-based models. This approach enables us to evaluate the performance of the proposed method utilizing self-supervised learning and spatial dependencies and temporal information. The strengths and improvement can be highlighted through comparing our model with existing state-of-the-art graph-based models, providing more efficient solutions to network intrusion detection. Descriptions of the existing graph-based methods are as follows.

E-GraphSAGE [9]: As the first application of GNN in NIDS, E-GraphSAGE is widely used for network intrusion detection task while obtaining the superior performance. It innovatively transforms network traffic data into a graph structure and utilizes the GNN model GraphSAGE and edge features in the network to learn a complex relationship between different hosts, which lay a solid foundation for the application of GNN in network intrusion detection.

**Table 8**
Comparison in binary classification results.

| Models | Datasets | Metrics | | | |
|---|---|---|---|---|---|
| | | Accuracy | Precision | Recall | F1-Score |
| E-GraphSAGE [9] | BoT-IoT | 99.99% | **1.00** | 99.99% | **1.00** |
| NEGAT+NEGSC [15] | BoT-IoT | 99.99% | 0.99 | 99.99% | 0.99 |
| TCG-IDS [19] | BoT-IoT | 99.45% | 1.00 | 99.45% | 0.99 |
| Ours | BoT-IoT | **99.99%** | 0.99 | **99.99%** | 0.99 |
| E-GraphSAGE | ToN-IoT | 97.87% | **1.00** | 97.87% | **1.00** |
| NEGAT+NEGSC | ToN-IoT | 96.76% | 0.96 | 99.81% | 0.98 |
| TCG-IDS | ToN-IoT | 97.54% | 0.95 | 97.54% | 0.95 |
| Ours | ToN-IoT | **98.57%** | 0.98 | **99.26%** | 0.99 |
| E-GraphSAGE | UNSW-NB15 | 98.76% | 0.91 | **99.99%** | 0.95 |
| NEGAT+NEGSC | UNSW-NB15 | 96.97% | 0.91 | 84.17% | 0.87 |
| TCG-IDS | UNSW-NB15 | 97.83% | 0.90 | 97.83% | 0.95 |
| Ours | UNSW-NB15 | **99.05%** | **0.95** | 96.31% | **0.97** |
| E-GraphSAGE | NF-UNSW-NB15-v3 | **99.98%** | 0.99 | **99.98%** | 0.99 |
| NEGAT+NEGSC | NF-UNSW-NB15-v3 | 98.61% | 0.98 | 75.74% | 0.85 |
| TCG-IDS | NF-UNSW-NB15-v3 | 98.87% | 0.98 | 98.87% | 0.96 |
| Ours | NF-UNSW-NB15-v3 | 99.95% | **0.99** | 99.54% | **0.99** |

NEGAT+NEGSC [15]: NEGAT+NEGSC is the first method that applied self-supervised learning and GNN in NIDS on both binary and multiclass classifications, getting the great results while reducing the reliance of labeled data. It improves the graph attention mechanism using edge features, named NEGAT, and proposes a self-supervised learning method based on graph contrastive learning, which considers the relationship between local and global topology of network, evaluating the potential of self-supervised learning in network intrusion detection.

TCG-IDS [19]: To further explore the advantages of the raw time in datasets, we compare it with TCG-IDS which utilized self-supervised learning and the generated time to implement the experiments. TCG-IDS proposes three contrastive learning strategies (e.g., temporal, asymmetric and masking) to capture the temporal dependencies, consider the different interactions within network data, and improve the ability to learn the representation, respectively, achieving superior performance in two different netflow datasets without timestamps (e.g., NF-CSE-CIC-IDS2018-v2 and NF-UNSW-NB15-v2).

The results of our method on binary classification indicates the outstanding performance on BoT-IoT, ToN-IoT, UNSW-NB15 and NF-UNSW-NB15-v3 datasets. On BoT-IoT, the proposed method has a accuracy of 99.99%, precision of 0.99, recall of 99.99% and F1-Score of 0.99, which achieves the superior performance of E-GraphSAGE and NEGAT+NEGSC and is slightly higher than TCG-IDS. On ToN-IoT, our method achieved a accuracy of 98.57%, precision of 0.98, recall of 99.26% and F1-Score of 0.99, outperforming E-GraphSAGE in accuracy and recall, and NEGAT+NEGSC and TCG-IDS in most metrics. The results of our method on UNSW-NB15 are accuracy of 99.05%, precision of 0.95, recall of 96.31% and F1-Score of 0.97, which are higher than two self-supervised learning methods and on a pair with the supervised E-GraphSAGE. Moreover, our method has a competitive performance in the netflow dataset NF-UNSW-NB15-v3, achieving a accuracy of 99.95%, precision of 0.99, recall of 99.54% and F1-Score of 0.99, which outperforms NEGAT+NEGSC and TCG-IDS in all metrics, and is slightly lower than E-GraphSAGE. The detailed results are illustrated as Table. 8

The comparisons with self-supervised GNN-based methods of binary classifications on four datasets can demonstrate that the spatial information and temporal dependencies in original data with timestamps play a paramount role in identify whether a network flow is benign. Meanwhile, our method achieves or even outperforms the performance supervised GNN-based model E-GraphSAGE and outperforms two self-supervised models NEGSC+NEGAT and TCG-IDS on most datasets, which can indicate the model using real timestamps can has a competitive performance without the labeled data.

On multiclass classification, our method achieved the superior performance, as shown in Table. 9. Results on BoT-IoT achieve a recall of 99.93% and F1-Score of 0.99, which outperform the performance of the NEGAT+NEGSC and TCG-IDS, 68.66% and 94.63% in Recall and 0.68 and 0.97 in F1-Score, respectively, and is slightly lower than E-GraphSAGE, demonstrating the advantage and influence of spatial information and temporal dependencies in identifying the specific attack types with the self-supervised learning. Results on ToN-IoT achieve a recall of 84.52% and F1-Score of 0.83 in Recall, which outperforms NEGAT+NEGSC, TCG-IDS and E-GraphSAGE in both metrics. The reason why the performance is in the range of over 80% in ToN-IoT dataset we conclude is that it contains diverse attack types and there is an extreme imbalance among different types. Results on UNSW-NB15 have a recall of 98.31%, and F1-Score of 0.98, respectively, which have a superior performance over other three methods. Results on NF-UNSW-NB15-v3 achieve a recall of 97.81%, and F1-Score of 0.97, respectively, which outperform the performance of both self-supervised GNN-based methods in Recall and F1-Score, and are very close to the supervised method E-GraphSAGE.

The results compared with self-supervised GNN-based methods of multiclass classifications on four datasets can show that our proposed method can have a great performance under an environment of diverse attacks. The results of comparing with NEGAT+NEGSC which is a static self-supervised method can indicate the advantage of spatial dependencies and temporal information in the dynamically evolving network. Also, the real timestamp of network can play an essential role in detecting normal and different malicious flows through comparing with TCG-IDS. Meanwhile, our method has a comparable performance with the supervised method E-GraphSAGE in total, which demonstrates the effectiveness of self-supervised learning for identifying attacks and reducing the reliance of labeled data.

## 4.5. Ablation Study

To concretize the strenghth of three modules of our method, ablation experiments are implemented on two datasets (e.g., ToN-IoT and NF-UNSW-NB15-v3) through removing each module. The modules contains temporal contrastive

**Table 9**
Comparison in multiclass classification results.

| Models | Datasets | Metrics | |
|---|---|---|---|
| | | Recall | F1-Score |
| E-GraphSAGE [9] | BoT-IoT | **99.99%** | **1.00** |
| NEGAT+NEGSC [15] | | 68.66% | 0.68 |
| TCG-IDS [19] | | 94.63% | 0.97 |
| Ours | | 99.93% | 0.99 |
| E-GraphSAGE | ToN-IoT | 82.78% | 0.82 |
| NEGAT+NEGSC | | 69.53% | 0.62 |
| TCG-IDS | | 79.53% | 0.79 |
| Ours | | **84.52%** | **0.83** |
| E-GraphSAGE | UNSW-NB15 | 95.86% | 0.96 |
| NEGAT+NEGSC | | 96.04% | 0.95 |
| TCG-IDS [19] | | 95.76% | 0.89 |
| Ours | | **98.31%** | **0.98** |
| E-GraphSAGE | NF-UNSW-NB15-v3 | 97.36% | 0.97 |
| NEGAT+NEGSC | | 95.82% | 0.94 |
| TCG-IDS | | 95.77% | 0.95 |
| Ours | | **97.81%** | **0.97** |

**Table 10**
Ablation results in binary classification of ToN-IoT.

| Method | Dataset | Metrics | | | |
|---|---|---|---|---|---|
| | | Accuracy | Precision | Recall | F1-Score |
| w/o Temporal | ToN-IoT | 97.07% | 0.97 | 98.50% | 0.99 |
| w/o Spatial | | 97.53% | 0.97 | 98.73% | 0.99 |
| w/o Feature | | 97.67% | 0.97 | 98.80% | 0.99 |
| Proposed | | **98.66%** | **0.98** | **99.31%** | **0.99** |

learning, spatial contrastive learning, and feature contrastive learning.

The binary results on ToN-IoT and NF-UNSW-NB15 are illustrated in Table 10 and Table 11, respectively. After removing temporal contrastive module, we can clearly see that there is a great drop for all metrics, which has an accuracy, precision, recall and F1-score of 97.07%, 0.97, 98.50% and 0.99, respectively. The method of removed spatial contrastive module shows the performance increases slightly than temporal, with accuracy, precision, recall and F1-score of 97.53%, 0.97, 98.73% and 0.99, respectively. The method of dropped feature contrastive module has an accuracy, precision, recall and F1-score of 97.67%, 0.97, 98.80% and 0.99, respectively. The method with three modules shows a superior performance, with an accuracy, precision, recall and F1-score of 98.66%, 0.98, 99.31% and 0.99, which outperforms other three methods.

It has a similar tendency in binary experiments of NF-UNSW-NB15-v3. There is a great drop for all metrics after removing temporal contrastive module, which has an accuracy, precision, recall and F1-score of 98.61%, 0.99, 85.36% and 0.74, respectively. The experiment of removed spatial contrastive module shows the performance has a slight increase than temporal, with accuracy, precision, recall and F1-score of 99.10%, 0.99, 90.97% and 0.83, respectively. After dropping feature contrastive module, the method has an accuracy, precision, recall and F1-score of 99.17%, 0.99, 91.80% and 0.85, respectively. The full method shows a superior performance, with an accuracy, precision, recall and F1-score of 99.95%, 0.99, 99.54% and 0.99, which outperforms other methods with removed module.

**Table 11**
Ablation results in binary classification of NF-UNSW-NB15-v3.

| Method | Dataset | Metrics | | | |
|---|---|---|---|---|---|
| | | Accuracy | Precision | Recall | F1-Score |
| w/o Temporal | NF-UNSW-NB15-v3 | 98.61% | 0.99 | 85.36% | 0.74 |
| w/o Spatial | | 99.10% | 0.99 | 90.97% | 0.83 |
| w/o Feature | | 99.17% | 0.99 | 91.80% | 0.85 |
| Proposed | | **99.95%** | **0.99** | **99.54%** | **0.99** |

**Table 12**
Ablation results in multiclass classification of ToN-IoT.

| Methods | Dataset | Metrics | |
|---|---|---|---|
| | | Recall | F1-Score |
| w/o Temporal | ToN-IoT | 82.01% | 0.80 |
| w/o Spatial | | 82.81% | 0.81 |
| w/o Feature | | 82.97% | 0.81 |
| Proposed | | **84.52%** | **0.83** |

**Table 13**
Ablation results in multiclass classification of NF-UNSW-NB15-v3.

| Methods | Dataset | Metrics | |
|---|---|---|---|
| | | Recall | F1-Score |
| w/o Temporal | NF-UNSW-NB15-v3 | 96.29% | 0.95 |
| w/o Spatial | | 96.74% | 0.96 |
| w/o Feature | | 97.31% | 0.96 |
| Propoesd | | **97.81%** | **0.97** |

The ablation results of the ToN-IoT dataset on multi-classification tasks are presented in Table 12. When the temporal module is removed, the model achieves a Recall of 82.01% and an F1-Score of 0.80. In contrast, removing the spatial module increases Recall to 82.81% and F1-Score to 0.81, slightly outperforming the w/o Temporal results on both metrics. Without the feature module, the model achieved a Recall of 82.97% and an F1-Score of 0.81. Its Recall metric was slightly higher than the previous two pruning settings, while its F1-Score remained consistent with w/o Spatial. For the Proposed method incorporating all modules, the model achieved Recall and F1-Score of 84.52% and 0.83, respectively.

Table 13 reports the ablation results on the NF-UNSW-NB15-v3 dataset under the multiclass classification setting. When the temporal module is removed, the model achieves a Recall of 96.29% and an F1-Score of 0.95. Removing the spatial module results in a higher Recall of 96.74% and an F1-Score of 0.96, exceeding the performance of the w/o Temporal configuration on both metrics. In the absence of the feature module, the Recall further increases to 97.31%, while the F1-Score remains at 0.96, which is comparable to the w/o Spatial setting. With all modules included, the Proposed method attains the highest Recall (97.81%) and F1-Score (0.97) among all configurations. Across both binary and multiclass classification tasks on the ToN-IoT and NF-UNSW-NB15-v3 datasets, consistent trends can be observed from the ablation results. Removing the temporal contrastive module leads to the most pronounced performance degradation in terms of Recall and F1-Score, indicating its dominant contribution among the three components. The exclusion of the spatial module also results in noticeable performance drops, although its impact is generally less severe than that of the temporal module. By comparison, removing the feature module causes relatively smaller performance degradation, suggesting a complementary role in representation enhancement. The results show that the temporal contrast module plays an importance role in improving the model's performance. Meanwhile, the best performance is consistently achieved when all three modules are jointly employed, highlighting the necessity of their collaborative integration.

**Table 14**
Testing Time on Four Datasets.

| Models | Datasets | Testing Time (s) |
|---|---|---|
| NEGAT+NEGSC [15] | BoT-IoT | 0.060 |
| TCG-IDS [19] | | 0.054 |
| Ours | | **0.032** |
| NEGAT+NEGSC | ToN-IoT | 0.115 |
| TCG-IDS | | 0.063 |
| Ours | | **0.032** |
| NEGAT+NEGSC | UNSW-NB15 | 0.550 |
| TCG-IDS | | 0.376 |
| Ours | | **0.077** |
| NEGAT+NEGSC | NF-UNSW-NB15-v3 | 0.045 |
| TCG-IDS | | 0.047 |
| Ours | | **0.043** |

### 4.6. Speed Testing Experiments

To verify the detection speed of using the simple and effective bidirectional LSTM plus E-GraphSAGE as the spatio-temporal encoder, we conducted tests on four datasets. From the table 14, the proposed model demonstrates robust stability in both testing efficiency across four different datasets. On the BoT-IoT and ToN-IoT datasets, the model achieves a consistent testing time of 0.032 seconds, indicating sustained high inference efficiency even under large-scale IoT traffic scenarios. On the UNSW-NB15 dataset, the testing time slightly increased to 0.077 s, primarily due to the dataset's feature dimensionality and complex traffic patterns. On the NF-UNSW-NB15-v3 dataset, the testing time decreased back to 0.043 s, further demonstrating the model's strong adaptability across varying data distributions.79

The model delivers fast testing speeds while maintaining robust detection performance. These results are consistent with our theoretical analysis of time complexity, futher validating its feasibility and practical value for real-time intrusion detection in IoT environments.

## 5. Conclusion

This paper proposes a self-supervised GNN-based framework for network intrusion detection systems. To the best of our knowledge, it is the first to take advantage of real timestamps to effectively identify the types of network flows in a self-supervised manner. By integrating timestamp-aware spatio-temporal graph contrastive learning with an efficient E-GraphSAGE and LSTM based encoder, the proposed method captures temporal dependencies and spatial relations in network traffic without introducing computationally expensive attention mechanisms. Moreover, theoretical time complexity analysis shows that explicitly modeling temporal dependencies, rather than relying on attention mechanisms, leads to a lower computational cost.

Extensive experiments on four datasets with dataset-provided timestamps, BoT-IoT, ToN-IoT, UNSW-NB15 and NF-UNSW-NB15-v3, demonstrate the high computational efficiency and good generalization of the proposed method. In particular, attacks with temporally repetitive and structurally consistent interaction patterns (e.g., DoS and DDoS) are classified with notably higher accuracy, further highlighting the effectiveness of leveraging real timestamps by the proposed timestamp-aware spatio-temporal contrastive learning framework. In addition, the experimental results and ablation studies consistently confirm the importance of dataset-provided timestamps for intrusion detection.

In the future, we will further refine our modeling of time-dependent behaviors through more fine-grained temporal analysis, to achieve improved robustness of intrusion detection systems in dynamic real-world scenarios.

## Declaration of competing interest

The authors declare that they have no known competing financial interests or personal relationships that could have appeared to influence the work reported in this paper.

## References

[1] N. Chaabouni, M. Mosbah, A. Zemmari, C. Sauvignac, P. Faruki, Network intrusion detection for iot security based on learning techniques, IEEE Communications Surveys & Tutorials 21 (3) (2019) 2671–2701.

[2] J. Fu, L. Wang, J. Ke, K. Yang, R. Yu, Tsids: Spatialtemporal fusion gating multilayer perceptron for network intrusion detection, Expert Systems with Applications 263 (2025) 125687.

[3] Z. Chen, H. Zou, T. Hu, X. Yuan, X. Fang, Y. Pan, J. Li, Hc-nids: Historical contextual information based network intrusion detection system in internet of things, Computers & Security 152 (2025) 104367.

[4] M. Zhong, M. Lin, C. Zhang, Z. Xu, A survey on graph neural networks for intrusion detection systems: Methods, trends and challenges, Computers & Security 141 (2024) 103821.

[5] A. Thakkar, R. Lohiya, A review on machine learning and deep learning perspectives of ids for iot: Recent updates, security issues, and challenges., Archives of computational methods in engineering 28 (4) (2021).

[6] A. L. Buczak, E. Guven, A survey of data mining and machine learning methods for cyber security intrusion detection, IEEE Communications surveys & tutorials 18 (2) (2015) 1153–1176.

[7] N. Shone, T. N. Ngoc, V. D. Phai, Q. Shi, A deep learning approach to network intrusion detection, IEEE Transactions on Emerging Topics in Computational Intelligence 2 (1) (2018) 41–50.

[8] P. Veličković, G. Cucurull, A. Casanova, A. Romero, P. Lio, Y. Bengio, Graph attention networks, arXiv preprint arXiv:1710.10903 (2017).

[9] W. W. Lo, S. Layeghy, M. Sarhan, M. Gallagher, M. Portmann, E-graphsage: A graph neural network based intrusion detection system for iot, arXiv preprint arXiv:2103.16329 (2021).

[10] G. Duan, H. Lv, H. Wang, G. Feng, Application of a dynamic line graph neural network for intrusion detection with semisupervised

learning, IEEE Transactions on Information Forensics and Security 18 (2022) 699–714.
[11] Y. Liu, M. Jin, S. Pan, C. Zhou, Y. Zheng, F. Xia, P. S. Yu, Graph self-supervised learning: A survey, IEEE Transactions on Knowledge and Data Engineering 35 (6) (2022) 5879–5900.
[12] W. Ju, Y. Wang, Y. Qin, Z. Mao, Z. Xiao, J. Luo, J. Yang, Y. Gu, D. Wang, Q. Long, et al., Towards graph contrastive learning: A survey and beyond, arXiv preprint arXiv:2405.11868 (2024).
[13] Q. Liu, H. Zhang, Y. Zhang, L. Fan, X. Jin, Ssa-gat: Graph-based self-supervised learning for network intrusion detection, in: International Conference on Artificial Neural Networks, Springer, 2024, pp. 476–491.
[14] E. Caville, W. W. Lo, S. Layeghy, M. Portmann, Anomal-e: A self-supervised network intrusion detection system based on graph neural networks, Knowledge-Based Systems 258 (2022) 110030.
[15] R. Xu, G. Wu, W. Wang, X. Gao, A. He, Z. Zhang, Applying self-supervised learning to network intrusion detection for network flows with graph neural network, Computer Networks 248 (2024) 110495.
[16] Z. A. Sahili, M. Awad, Spatio-temporal graph neural networks: A survey, arXiv preprint arXiv:2301.10569 (2023).
[17] L. Van Langendonck, Ppt-gnn: a practical pre-trained temporal graph neural network for intrusion detection, Master's thesis, Universitat Politècnica de Catalunya (2024).
[18] E. S. Escriche, J. Nyberg, Y. Kim, G. Dán, Channel-centric spatio-temporal graph networks for network-based intrusion detection, in: 2024 IEEE Conference on Communications and Network Security (CNS), IEEE, 2024, pp. 1–9.
[19] C. Wu, J. Sun, J. Chen, M. Alazab, Y. Liu, Y. Xiang, TCG-IDS : Robust network intrusion detection via temporal contrastive graph learning, IEEE Transactions on Information Forensics and Security 20 (2025) 1475–1486.
[20] C.-H. Lee, Y.-Y. Su, Y.-C. Lin, S.-J. Lee, Machine learning based network intrusion detection, in: 2017 2nd IEEE International conference on computational intelligence and applications (ICCIA), IEEE, 2017, pp. 79–83.
[21] R. Vinayakumar, K. P. Soman, P. Poornachandran, Applying convolutional neural network for network intrusion detection, in: 2017 International conference on advances in computing, communications and informatics (ICACCI), IEEE, 2017, pp. 1222–1228.
[22] H. He, X. Sun, H. He, G. Zhao, L. He, J. Ren, A novel multimodal-sequential approach based on multi-view features for network intrusion detection, IEEE Access 7 (2019) 183207–183221.
[23] X. Liu, F. Zhang, Z. Hou, L. Mian, Z. Wang, J. Zhang, J. Tang, Self-supervised learning: Generative or contrastive, IEEE Transactions on Knowledge and Data Engineering 35 (1) (2021) 857–876.
[24] B. Perozzi, R. Al-Rfou, S. Skiena, Deepwalk: Online learning of social representations, in: Proceedings of the 20th ACM SIGKDD international conference on Knowledge discovery and data mining, 2014, pp. 701–710.
[25] A. Grover, J. Leskovec, node2vec: Scalable feature learning for networks, in: Proceedings of the 22nd ACM SIGKDD international conference on Knowledge discovery and data mining, 2016, pp. 855–864.
[26] T. N. Kipf, M. Welling, Variational graph auto-encoders, arXiv preprint arXiv:1611.07308 (2016).
[27] P. Veličković, W. Fedus, W. L. Hamilton, P. Liò, Y. Bengio, R. D. Hjelm, Deep graph infomax, arXiv preprint arXiv:1809.10341 (2018).
[28] Y. You, T. Chen, Y. Sui, T. Chen, Z. Wang, Y. Shen, Graph contrastive learning with augmentations, Advances in neural information processing systems 33 (2020) 5812–5823.
[29] Y. Han, L. Hui, H. Jiang, J. Qian, J. Xie, Generative subgraph contrast for self-supervised graph representation learning, in: European Conference on Computer Vision, Springer, 2022, pp. 91–107.
[30] E. Rossi, B. Chamberlain, F. Frasca, D. Eynard, F. Monti, M. Bronstein, Temporal graph networks for deep learning on dynamic graphs, arXiv preprint arXiv:2006.10637 (2020).
[31] K.-J. Chen, L. Liu, L. Jiang, J. Chen, Self-supervised dynamic graph representation learning via temporal subgraph contrast, ACM Transactions on Knowledge Discovery from Data 18 (1) (2023) 1–20.
[32] H. Wang, X. Di, Y. Wang, B. Ren, G. Gao, J. Deng, An intelligent digital twin method based on spatio-temporal feature fusion for iot attack behavior identification, IEEE Journal on Selected Areas in Communications 41 (11) (2023) 3561–3572.
[33] B. Yu, H. Yin, Z. Zhu, Spatio-temporal graph convolutional networks: A deep learning framework for traffic forecasting, arXiv preprint arXiv:1709.04875 (2017).
[34] G. Duan, H. Lv, H. Wang, G. Feng, X. Li, Practical cyber attack detection with continuous temporal graph in dynamic network system, IEEE Transactions on Information Forensics and Security 19 (2024) 4851–4864.
[35] Z. Lu, W. Wang, C. Wang, On the evolution and impact of mobile botnets in wireless networks, IEEE Transactions on Mobile Computing 15 (9) (2016) 2304–2316.
[36] N. Samia, S. Saha, A. Haque, Predicting and mitigating cyber threats through data mining and machine learning, Computer Communications 228 (2024) 107949.
[37] X. Zang, J. Gong, M. Wang, P. Gao, G. Zhang, Ip traffic behavior characterization via semantic mining, Journal of Network and Computer Applications 213 (2023) 103603.
[38] X. Ma, J. Wu, S. Xue, J. Yang, C. Zhou, Q. Z. Sheng, H. Xiong, L. Akoglu, A comprehensive survey on graph anomaly detection with deep learning, IEEE transactions on knowledge and data engineering 35 (12) (2021) 12012–12038.
[39] A. Paszke, S. Gross, F. Massa, A. Lerer, J. Bradbury, G. Chanan, T. Killeen, Z. Lin, N. Gimelshein, L. Antiga, et al., Pytorch: An imperative style, high-performance deep learning library, Advances in neural information processing systems 32 (2019).
[40] M. Fey, J. E. Lenssen, Fast graph representation learning with pytorch geometric, arXiv preprint arXiv:1903.02428 (2019).
[41] N. Koroniotis, N. Moustafa, E. Sitnikova, B. Turnbull, Towards the development of realistic botnet dataset in the internet of things for network forensic analytics: Bot-iot dataset, Future Generation Computer Systems 100 (2019) 779–796.
[42] N. Moustafa, TON-IoT Dataset, [Online], available: https://cloudstor.aarnet.edu.au/plus/s/ds5zW91vdgjEj9i (2020).
[43] N. Moustafa, J. Slay, Unsw-nb15: a comprehensive data set for network intrusion detection systems (unsw-nb15 network data set), in: 2015 military communications and information systems conference (MilCIS), IEEE, 2015, pp. 1–6.
[44] M. Luay, S. Layeghy, S. Hosseininoorbin, M. Sarhan, N. Moustafa, M. Portmann, Temporal analysis of netflow datasets for network intrusion detection systems, arXiv preprint arXiv:2503.04404 (2025).